\newif\ifcol
\coltrue

\newif\ifbw
\bwfalse

\newif\ifdraft
\draftfalse

\newcommand{\pagesize}{letter}

\ifdraft
  \colfalse
\fi

\newlength{\mapwidth}
\ifdraft
\fi

\newlength{\coeffwidth}
\ifdraft
  \documentclass[11pt,draftcls,\pagesize,onecolumn,twoside]{IEEEtran}
\else
  \documentclass[10pt,journal,\pagesize,twoside]{IEEEtran}
\fi

\usepackage{url}

\usepackage{ifpdf}
\ifpdf
  \usepackage[pdftex,
    pdfauthor     = {Jason McEwen},
    pdftitle      = {Fast directional CSWT algorithms},
    pdfsubject    = {Continuous spherical wavelets},
    pdfkeywords   = {Wavelet transforms -- spheres -- convolution},
    pdfproducer   = {pdflatex with hyperref},
    pdfcreator    = {pdflatex}]{hyperref}
\else
  \usepackage[ps2pdf,colorlinks=true]{hyperref}
  \hypersetup{%
    pdfauthor     = {Jason McEwen},
    pdftitle      = {Fast directional CSWT algorithms},
    pdfsubject    = {Continuous spherical wavelets},
    pdfkeywords   = {Wavelet transforms -- spheres -- convolution},
    pdfauthor     = {ps2pdf with hyperref},
    pdfcreator    = {LateX},
  }
\fi

\usepackage{cite}      

\ifpdf
  \usepackage[pdftex]{graphicx}
\else
  \usepackage[dvips]{graphicx}
\fi

\usepackage{subfigure} 

\usepackage{amssymb}    

\newcommand{\cswtfftterm}{\ensuremath{T}}

\newcommand{\el}{\ensuremath{\ell}}
\newcommand{\m}{\ensuremath{m}}
\newcommand{\elm}{\ensuremath{{\el \m}}}
\newcommand{\mmax}{\ensuremath{\m_{\rm max}}}
\newcommand{\elmax}{\ensuremath{{\el_{\rm max}}}}
\newcommand{\dmatbig}{\ensuremath{D}}
\newcommand{\dmatsmall}{\ensuremath{d}}
\newcommand{\admissc}{\ensuremath{C}}
\newcommand{\sh}{\ensuremath{Y}}
\newcommand{\shcoeff}[1]{\ensuremath{\widehat{#1}}}

\newcommand{\aleg}[3]{\ensuremath{P_{#1}^{#2}({#3})}}

\newcommand{\eqn}[1]
	{(#1)}

\newcommand{\tbl}[1]
	{Table~#1}

\newcommand{\fig}[1]
	{Fig.~#1}

\newcommand{\sectn}[1]
	{section~#1}

\newcommand{\eg}{\mbox{\it{e.g.}}}
\newcommand{\ie}{\mbox{\it{i.e.}}}

\newcommand{\fft}{{FFT}}
\newcommand{\dft}{{DFT}}

\newcommand{\isw}
	{{ISW}}
\newcommand{\iswtext}
	{integrated Sachs-Wolfe}

\newcommand{\cswt}
	{{CSWT}}

\newcommand{\smhw}{{SMHW}}
\newcommand{\smw}{{SMW}}
\newcommand{\sbw}{{SBW}}

\newcommand{\yawtb}{{YAWTb}}
\newcommand{\healpix}{{HEALPix}}

\newcommand{\xvect}{\ensuremath{\mathbf{x}}}
\newcommand{\kvect}{\ensuremath{\mathbf{k}}}

\newcommand{\sky}{\ensuremath{s}}
\newcommand{\skywav}{\ensuremath{{W_\wav^\sky}}}

\newcommand{\spcend}{\ensuremath{\:}}

\newcommand{\img}{\ensuremath{\mathit{i}}}
\newcommand{\dx}{\ensuremath{\mathrm{\,d}}}
\newcommand{\dmu}{\ensuremath{\dx \Omega}}

\newcommand{\cconj}{\ensuremath{\ast}}  
\newcommand{\sa}{\ensuremath{\omega}}
\newcommand{\saa}{\ensuremath{\theta}}
\newcommand{\sab}{\ensuremath{\phi}}
\newcommand{\sas}{\ensuremath{\saa, \sab}}
\newcommand{\eul}{\ensuremath{\mathbf{\rho}}}
\newcommand{\euls}{\ensuremath{\eula, \eulb, \eulc}}
\newcommand{\eula}{\ensuremath{\alpha}}
\newcommand{\eulb}{\ensuremath{\beta}}
\newcommand{\eulc}{\ensuremath{\gamma}}

\newcommand{\spo}{\ensuremath{\Pi}}

\newcommand{\cocycle}{\ensuremath{\lambda}}

\newcommand{\realno}{\ensuremath{\mathbb{R}}}
\newcommand{\natno}{\ensuremath{\mathbb{N}}}
\newcommand{\sothree}{\ensuremath{\mathrm{SO}(3)}}
\newcommand{\sphere}{\ensuremath{{S^2}}}
\newcommand{\dil}{\ensuremath{\mathcal{D}}}
\newcommand{\dilsmall}{\ensuremath{d}}
\newcommand{\rot}{\ensuremath{R}}
\newcommand{\scalea}{\ensuremath{a}}
\newcommand{\scaleb}{\ensuremath{b}}
\newcommand{\scaleab}{\ensuremath{{\scalea,\scaleb}}}

\newcommand{\rad}{\ensuremath{r}}
\newcommand{\p}{\ensuremath{^\prime}}
\newcommand{\pp}{\ensuremath{^{\prime\prime}}}
\newcommand{\wav}{\ensuremath{\psi}}

\newcommand{\grideula}{\ensuremath{\mathcal{E}_1}}
\newcommand{\grideulb}{\ensuremath{\mathcal{E}_2}}
\newcommand{\gridhpix}{\ensuremath{\mathcal{H}}}
\newcommand{\gridecp}{\ensuremath{\mathcal{C}}}
\newcommand{\ia}{\ensuremath{{n_\eula}}}
\newcommand{\ib}{\ensuremath{{n_\eulb}}}
\newcommand{\ig}{\ensuremath{{n_\eulc}}}
\newcommand{\ik}{\ensuremath{{k}}}
\newcommand{\ith}{\ensuremath{{n_\saa}}}
\newcommand{\iph}{\ensuremath{{n_\sab}}}
\newcommand{\ipix}{\ensuremath{{p}}}
\newcommand{\na}{\ensuremath{{N_\eula}}}
\newcommand{\nb}{\ensuremath{{N_\eulb}}}
\newcommand{\ngm}{\ensuremath{{N_\eulc}}}
\newcommand{\nth}{\ensuremath{{N_\saa}}}
\newcommand{\nph}{\ensuremath{{N_\sab}}}
\newcommand{\n}{\ensuremath{{N}}}
\newcommand{\nside}{\ensuremath{{N_{\rm{side}}}}}
\newcommand{\npix}{\ensuremath{{N_{\rm{pix}}}}}
\newcommand{\weight}{\ensuremath{w}}
\newcommand{\order}{\ensuremath{\mathcal{O}}}

\newcommand{\etal}{\mbox{\it{et al.}}}

\begin{document}
\title{Fast directional continuous spherical wavelet transform algorithms}
%
%
\author{J.~D.~McEwen,~M.~P.~Hobson,~D.~J.~Mortlock,~A.~N.~Lasenby%
\thanks{Manuscript received 14 June, 2005}
\thanks{J.~D.~McEwen,~M.~P.~Hobson and A.~N.~Lasenby are with the 
        Astrophysics Group, Cavendish Laboratory, Cambridge, UK.}
\thanks{D.~J.~Mortlock is with the Institute of Astronomy, Cambridge, UK.}
\thanks{E-mail: mcewen@mrao.cam.ac.uk (J.~D.~McEwen)}}
%
%
%
\markboth{IEEE Transactions on Signal Processing,~Vol.~--, No.~--,~June~2005}%
{McEwen \MakeLowercase{\textit{et al.}}: Fast directional \cswt\ algorithms}
%



\maketitle



\begin{abstract}
We describe the construction of a spherical wavelet analysis through
the inverse stereographic projection of the Euclidean planar wavelet
framework, introduced originally by Antoine and Vandergheynst and developed further by
Wiaux \etal.  
Fast algorithms for performing the directional continuous wavelet
analysis on the unit sphere are presented.  The fast directional algorithm,
based on the fast spherical convolution algorithm developed by Wandelt
and G\'{o}rski,
provides a saving of $\order(\sqrt{\npix})$ over a direct quadrature
implementation for \npix\ pixels on the sphere, and 
allows one to perform a directional spherical wavelet analysis of a 
$10^6$ pixel map on a personal computer.
\end{abstract}

\begin{keywords}
Wavelet transforms, spheres, convolution.
\end{keywords}

\IEEEpeerreviewmaketitle

\section{Introduction}
\label{sec:intro}

\PARstart{W}{avelet} analysis has proven useful in many
applications due to 
the ability of wavelets to resolve localised signal content in both
scale and space.  Many of these applications, however, are restricted
to data defined in Euclidean space: the 1-dimensional line, 
the 2-dimensional plane and,
occasionally, higher dimensions.  Nevertheless, data are often measured
or defined on other manifolds, such as the 2-sphere.  
Examples where data are measured on the sphere are found in 
astrophysics (\eg\ \cite{bennett:1996,bennett:2003}), 
planetary science (\eg\ \cite{wieczorek:2006,wieczorek:1998,turcotte:1981}), 
geophysics (\eg\ \cite{whaler:1994,swenson:2002,simons:2006}), 
computer vision (\eg\ \cite{ramamoorthi:2004}) and 
quantum chemistry (\eg\ \cite{choi:1999,ritchie:1999}).
%
To realise the potential benefits that may be provided by wavelets
in such settings, ordinary Euclidean wavelet analysis must be extended
to spherical geometry.

A number of attempts have been made to extend wavelets to the unit sphere.  
Discrete second generation wavelets on the sphere that are based on a multiresolution analysis have been developed \cite{schroder:1995,sweldens:1996}.  Other authors have focused on the continuous wavelet transform on the sphere.  A number of works construct a solution using a harmonic approach \cite{narcowich:1996,potts:1996,freeden:1997a,freeden:1997b}, however these solutions suffer from the poor localisation of the spherical harmonic functions.  Others adopt a tangent bundle viewpoint \cite{torresani:1995,dahlke:1996}, thereby avoiding the necessity to define a dilation operator on the sphere.  A satisfactory extension of the continuous wavelet transform to the sphere is defined by \cite{holschneider:1996}, however this construction requires an abstract dilation parameter that must satisfy a number of {\it ad hoc} assumptions.
%
%
More recently, a consistent and
satisfactory framework for wavelets defined on the unit sphere has been
constructed and developed by 
\cite{antoine:1999,antoine:1998,antoine:2002,antoine:2004,%
bogdanova:2004,demanet:2003,wiaux:2005}.  Moreover, this construction is derived entirely from group theoretic principles and inherently satisfies a number of natural requirements.
We consider the continuous spherical wavelet transform (\cswt)
developed in these last works.  
For a more detailed review of the attempts made at constructing a wavelet transform on the unit sphere see \cite{antoine:1999,antoine:1998,mhaskar:2001}.

Current and future data-sets defined on the sphere are of
considerable size.  The current Wilkinson Microwave Anisotropy Probe (WMAP) data of the cosmic microwave
background (CMB) contain approximately $3\times10^6$ pixels on the
sphere, whereas the forthcoming Planck CMB mission will generate maps with  
approximately $50\times10^6$ pixels.  
Fast algorithms are therefore required to perform the \cswt\ on
practical data-sets.
A semi-fast algorithm to implement the \cswt\ is presented in
\cite{antoine:2002} and implemented in the 
\yawtb\footnote{\url{http://www.fyma.ucl.ac.be/projects/yawtb/}}
(Yet-Another-Wavelet-Toolbox) Matlab wavelet
toolbox (which also makes use of the SpharmonicKit\footnote{\url{http://www.cs.dartmouth.edu/~geelong/sphere/}}).
However, this implementation is restricted to an equi-angular
tessellation of the sphere.  The beauty of this tessellation is its
simplicity and ability to be easily represented in matrix form.
However, the pixels of an equi-angular tessellation are densely spaced
about the poles and do not have equal areas. 
Other tessellations of the sphere also exist, such as those constructed to minimise some energy measure \cite{saff:1997,hardin:2004,du:2003} or those constructed for more practical or numerical purposes (for example
the  IGLOO\footnote{\url{http://www.mrao.cam.ac.uk/projects/cpac/igloo/}}
\cite{crittenden:1998},
\healpix\footnote{\url{http://healpix.jpl.nasa.gov/}} 
\cite{gorski:2005}
and GLESP\footnote{\url{http://www.glesp.nbi.dk/}} \cite{doroshkevich:2005}
schemes).
There is thus a need for a fast implementation of the
\cswt\ that is 
not tied to any particular tessellation of the sphere.

We fill this void by presenting a fast algorithm for performing
the directional \cswt. 
The \cswt\ at a particular scale is essentially a
spherical convolution, hence we may apply the fast spherical convolution
algorithm developed
by \cite{wandelt:2001} to evaluate the wavelet transform.
The algorithm is posed in harmonic space and thus is
independent of the underlying tessellation of the sphere, (although an
iso-latitude tessellation does enable faster spherical harmonic
transforms, thereby increasing the speed of the algorithm). 
The framework supports both non-azimuthally symmetric spherical wavelets\footnote{Azimuthally symmetric spherical wavelets are also often referred to as axisymmetric wavelets.} and a decomposition that employs anisotropic dilations, however no synthesis is possible when anisotropic dilations are incorporated.
For an illustration of a spherical wavelet analysis on a practical
problem of considerable size we refer the reader to our recent works to
test the CMB for deviations from Gaussianity \cite{mcewen:2004,mcewen:2006b} and to detect the \iswtext\ (\isw) effect \cite{mcewen:2006}.
Both of these works involve performing 1000 Monte Carlo simulations and  
would not have been feasible without a fast directional \cswt\ algorithm.

The remainder of this paper is structured as follows.  
In \sectn{\ref{sec:cswt}} we describe the \cswt\ in the framework
presented by \cite{wiaux:2005}.  
For a more complete treatment of the spherical wavelet transform in
this framework and the correspondence between spherical and Euclidean
wavelets we recommend that the reader refer to \cite{wiaux:2005}.
We also present an extension to anisotropic dilations, however in this case the basis functions are not strictly wavelets hence perfect reconstruction is not possible.
%
Various algorithms to perform the \cswt\ are
described and then compared in \sectn{\ref{sec:algorithms}}.
An application of our implementation is demonstrated in \sectn{\ref{sec:application}}.
Concluding remarks are made in \sectn{\ref{sec:conclusions}}.

\section{The continuous spherical wavelet transform}
\label{sec:cswt}

To extend Euclidean wavelet analysis to spherical geometry a number of
requirements must be satisfied: (i) the signals and wavelets
must live fully on the unit sphere; (ii) the transform must involve local
dilations of some kind on the unit sphere; and (iii) the spherical wavelet
transform  should reduce locally to the Euclidean transform on the
tangent plane (\ie\ the Euclidean limit must be satisfied)
\cite{holschneider:1996}.
The final requirement is intuitively obvious; the sphere is asymptotically
flat, hence any spherical wavelet transform should match the planar
Euclidean transform on small scales, or equivalently, for a large
radius of curvature.

The spherical wavelet transform developed in
\cite{antoine:1999,antoine:1998,antoine:2002,antoine:2004,bogdanova:2004,demanet:2003}
satisfies all of these requirements, moreover each requirement
naturally follows from the construction.  The construction of this
transform is derived entirely from group theoretical principles.
However, in a recent work by \cite{wiaux:2005} this formalism is
reintroduced independently of the original group theoretic
formalism, in an equivalent, practical and self-consistent approach. 
We adopt this approach herein.

The correspondence principle between spherical and Euclidean wavelets
is developed by \cite{wiaux:2005}, relating the concepts of planar
Euclidean wavelets to spherical wavelets through a stereographic
projection.  We use the stereographic projection to define affine
transformations on the unit sphere that facilitate the construction of a
wavelet basis on the unit sphere.  The spherical wavelet transform may then
be defined as the projection on to this basis, where the spherical
wavelets must satisfy the appropriate admissibility criterion to
ensure perfect reconstruction.

\subsection{Stereographic projection}

In order to construct a correspondence between wavelets on the plane
($\realno^2$) and sphere (\sphere) a projection operator between the
two spaces must be chosen.  
It is shown in \cite{wiaux:2005} that the stereographic projection is the unique unitary, radial and conformal diffeomorphism between the sphere and the plane.

\begin{figure}
\centerline{
  \includegraphics[width=86mm]{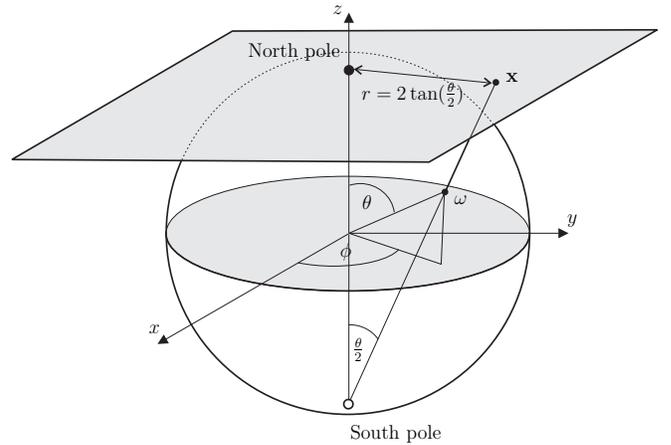}}
\caption[Stereographic projection]{Stereographic projection of the
  sphere onto the plane.}
\label{fig:stereographic_projection}
\end{figure}

The stereographic projection is defined by projecting a point on the
unit sphere to a point on the tangent plane at the north pole, by casting a
ray though the point and the south pole.  The point on the unit sphere is
mapped on to the intersection of this ray and the tangent plane (see
\fig{\ref{fig:stereographic_projection}}).  
Formally, we may define the stereographic projection operator as 
$\spo : \sa \rightarrow \xvect = \spo \sa = (\rad(\saa),\sab)$ where 
$\rad=2 \tan(\saa/2)$, $\sa \equiv (\sas) \in \sphere$ denotes
spherical coordinates with colatitude \saa\ and longitude \sab\
and $\xvect \in \realno^2$ is a point in the plane, denoted here by
the polar coordinates $(\rad,\sab)$.
The inverse operator is 
$\spo^{-1} : \xvect \rightarrow \sa = \spo^{-1} \xvect =
(\saa(\rad),\sab)$,
where $\saa(\rad)=2\tan^{-1}(\rad/2)$.

Following the formulation of \cite{wiaux:2005} again, we define the
action of the stereographic projection operator on functions on the
plane and sphere.  We consider the space of square integrable
functions in $L^2(\realno^2,\dx^2\xvect)$ on the plane and 
$L^2(\sphere,\dmu)$ on the unit sphere, where 
$\dmu = \sin\saa \dx\saa \dx\sab$ is the usual rotation invariant measure
on the sphere. 
The action of the stereographic projection operator 
$\spo : s \in L^2(\sphere,\dmu) \rightarrow p=\spo s \in L^2(\realno^2,\dx^2\xvect)$
on functions is defined as
\begin{equation}
\label{eqn:sp}
p(\rad,\sab) = (\spo s)(\rad,\sab) =
(1 + r^2/4)^{-1}
s(\saa(\rad),\sab)
\spcend .
\end{equation}
The inverse stereographic projection operator 
$\spo^{-1} : p \in L^2(\realno^2,\dx^2\xvect) \rightarrow s=\spo^{-1}
p \in L^2(\sphere,\dmu)$
on functions is then
\begin{equation}
\label{eqn:isp}
s(\sas) = (\spo^{-1} p)(\sas) =
[1 + \tan^2(\saa/2)]
p(\rad(\saa),\sab)
\spcend .
\end{equation}
The pre-factors introduced ensure that the $L^2$-norm of functions
through the forward and inverse projections are conserved. 
In the Euclidean limit, the stereographic projection and inverse
naturally reduce to the identity operator \cite{antoine:1999}. 

\subsection{Affine transformations on the sphere}
\label{sec:affine_transformations}

A wavelet basis is constructed on the unit sphere in
\sectn{\ref{sec:wavelet_transform}} by applying the 
spherical extension of Euclidean translations and dilations 
to mother wavelets defined on the unit sphere.  The extension of these
affine transformations to the sphere, facilitated by the stereographic
projection operator, are defined here.

The natural extension of Euclidean translations on the unit sphere are rotations.
These are characterised by the elements of the rotation group \sothree,
which we parameterise in terms of the three Euler angles
$\rho=(\euls)$.\footnote{We adopt the $zyz$ Euler convention corresponding to the rotation of a physical body in a \emph{fixed} co-ordinate system about the $z$, $y$ and $z$ axes by \eulc, \eulb\ and \eula\ respectively.}  The rotation of a square-integrable function 
$s\in L^2(\sphere,\dmu)$ is defined by
\begin{equation}
[\rot(\rho) s](\sa) = s(\rho^{-1} \sa), \; \; \rho \in \sothree 
\spcend .
\end{equation}

Dilations on the unit sphere are constructed by first
lifting the sphere to the plane by the stereographic projection,
\mbox{followed} by the usual Euclidean dilation in the plane, before
re-projecting back onto the sphere.  We generalise to anisotropic dilations on the sphere (a similar anisotropic dilation operator on the sphere has been independently proposed by \cite{tosic:2005}), however in this setting we do not achieve a wavelet basis and hence cannot synthesise our original signal.
We define the anisotropic
Euclidean dilation operator in $L^2(\realno^2,\dx^2\xvect)$ as
\begin{equation}
[\dilsmall(\scalea,\scaleb) p] (x,y) = \scalea^{-1/2}\scaleb^{-1/2} \:
p(\scalea^{-1}x,\scaleb^{-1}y)
\spcend ,
\end{equation}
for the non-zero positive scales 
$\scalea, \scaleb \in \realno^{+}_{\ast}$.
The $\scalea^{-1/2}\scaleb^{-1/2}$ normalisation factor ensures the
$L^2$-norm is preserved.  The spherical dilation operator 
$\dil(\scaleab) : s(\sas) \rightarrow [\dil(\scaleab)s](\sas)$
in $L^2(\sphere,\dmu)$ is defined as the conjugation by \spo\ of
the Euclidean dilation $d(\scaleab)$ in $L^2(\realno^2,\dx^2\xvect)$
on the tangent plane at the north pole:
\begin{equation}
\dil(\scaleab) = \spo^{-1} \, \dilsmall(\scaleab) \, \spo
\spcend .
\end{equation}
The norm of functions in $L^2(\sphere,\dmu)$ is preserved by the
spherical dilation as both the stereographic projection operator and
Euclidean dilations preserve the norm of functions.
Extending the isotropic spherical dilation operator defined by 
\cite{wiaux:2005} to anisotropic dilations, we obtain
\begin{equation}
[\dil(\scaleab) s](\sa)
= [\cocycle(\scaleab,\sas)]^{1/2} \: s(\sa_{1/\scalea,1/\scaleb})
\spcend ,
\end{equation}
where $\sa_\scaleab=(\saa_\scaleab,\sab_\scaleab)$,
\begin{displaymath}
\tan(\saa_\scaleab/2) = 
\tan(\saa/2)
\sqrt{\scalea^2 \cos^2{\sab} + \scaleb^2 \sin^2{\sab}}
\end{displaymath}
and 
\begin{displaymath}
\tan(\sab_\scaleab) = 
\frac{\scaleb}{\scalea}
\tan(\sab)
\spcend .
\end{displaymath}
For the case where $\scalea=\scaleb$ the anisotropic
dilation reduces to the usual isotropic case defined by 
$\tan(\saa_\scalea/2) = \scalea \tan(\saa/2)$ and
$\sab_\scalea=\sab$.  
The $\cocycle(\scaleab,\sas)$ cocycle term follows from the factors
introduced in the stereographic projection of functions to preserve
the $L^2$-norm.  Alternatively, the cocycle may be derived
explicitly to preserve the $L^2$-norm when the stereographic projection
of functions do not have these pre-factor terms.  The cocycle of an
anisotropic spherical dilation is defined by
\begin{equation}
\cocycle(\scaleab, \sas) =
\frac{4 \scalea^3 \scaleb^3}
{ 
\left(
A_{-}\cos\saa + A_{+}
\right)^2
}
\spcend ,
\end{equation}
where 
\begin{displaymath}
A_\pm = \scalea^2\scaleb^2 \pm \scalea^2 \sin^2\sab \pm \scaleb^2 \cos^2\sab
\spcend .
\end{displaymath}
For the case where $\scalea=\scaleb$ the anisotropic cocycle reduces
to the usual isotropic cocycle
\begin{displaymath}
\cocycle(\scalea, \scalea, \sas) =
\frac{4 \scalea^2}
{ [(\scalea^2-1)\cos{\saa} + \scalea^2+1]^2 }
\spcend .
\end{displaymath}
Although the ability to perform anisotropic dilations is of practical use, we do not achieve a wavelet basis in this setting. 
In the anisotropic setting a bounded admissibility integral cannot be determined (even in the plane), thus the synthesis of a signal from its coefficients cannot be performed. 
This results from there being no direct means of evaluating the proper measure in the absence of a group structure.
The projection of a signal onto basis functions undergoing anisotropic dilations may be performed in an analogous manner to the following discussion of the wavelet transform.  However, since these basis functions are not wavelets we restrict the following discussion to isotropic dilations.

\subsection{Wavelet transform}
\label{sec:wavelet_transform}

A wavelet basis on the unit sphere may now be constructed from rotations and
isotropic dilations (where $\scalea=\scaleb$) of a mother spherical wavelet $\wav \in L^2(\sphere,\dmu)$. 
The corresponding wavelet family 
\mbox{$\{ \wav_{\scalea,\rho} \equiv \rot(\eul) \dil(\scalea,\scalea) \wav,
\; {\rho \in \sothree,} \; {\scalea \in \realno_{\ast}^{+}} \}$}
provides an over-complete set of functions in $L^2(\sphere,\dmu)$. 
The \cswt\ of $\sky \in L^2(\sphere,\dmu)$ is given
by the projection onto each wavelet basis function in the usual
manner,
\begin{equation}
\skywav(\scalea, \eul) 
\equiv
\int_{\sphere}
\dmu \:
\wav_{\scalea,\eul}^\cconj(\sa) \:
\sky(\sa)
=
\left< \wav_{\scalea,\eul} \mid \sky \right>
\spcend ,
\label{eqn:cswt}
\end{equation}
where the \cconj\ denotes complex conjugation.

The transform is general in the sense that all orientations in the
rotation group \sothree\ are considered, thus directional structure is
naturally incorporated.  It is important to note, however, that only
\emph{local} directions make any sense on \sphere.  There is no global
way of defining directions on the sphere\footnote{There is no
differentiable vector field of constant norm on the sphere and hence
no global way of defining directions.} -- there will always be some
singular point where the definition fails.  

The synthesis of a signal on the unit sphere from its wavelet coefficients
is given by
  \begin{equation}
  \sky(\sa) = 
  \int_{\sothree} \dx\eul
  \int_0^\infty \frac{\dx\scalea}{\scalea^3} \:\:
  \skywav(\scalea, \eul) \:
  [\rot(\eul) L_\wav \wav_\scalea](\sa)
  \spcend ,
  \end{equation}
where $\dx\eul=\sin\eulb \dx\eula\dx\eulb\dx\eulc$.
The $L_\wav$ operator in $L^2(\sphere,\dmu)$ is defined by the
action
\begin{equation}
(\shcoeff{L_\wav f})_\elm = \shcoeff{f}_\elm / \admissc_\wav^\el 
\end{equation}
on the spherical harmonic coefficients of functions \mbox{$f \in L^2(\sphere,\dmu)$}, 
where  $\admissc_\wav^\el$ is defined below.
The hat denotes the spherical harmonic coefficients 
\begin{displaymath}
\shcoeff{f}_\elm 
= 
\int_{\sphere}
\dmu \:
\sh_\elm^\cconj(\sa) \:
f(\sa)
= \left< \sh_\elm \mid f \right> 
\end{displaymath}
of the decomposition
\begin{displaymath}
f(\sa) = \sum_{\el=0}^\infty \sum_{\m=-\el}^\el \shcoeff{f}_\elm \sh_\elm(\sa)
\spcend .
\end{displaymath}
We adopt the Condon-Shortley phase convention where the normalised 
spherical harmonics are defined by
\begin{displaymath}
\sh_{\elm}(\sa) = (-1)^\m \sqrt{\frac{2\el+1}{4\pi} 
\frac{(\el-\m)!}{(\el+\m)!}} \: 
\aleg{\el}{\m}{\cos\saa} \:
{ e}^{\img \m \sab}
\spcend ,
\end{displaymath}
where $\aleg{\el}{\m}{x}$ are the associated Legendre functions.  Using this normalisation the orthogonality of the spherical harmonic functions is given by
\begin{equation}
\label{eqn:shortho}
\int_\sphere
\dmu \:
\sh_{\elm}(\sa)
\sh_{\el\p \m\p}^\cconj(\sa)
=
\delta_{\el\el\p} \delta_{\m\m\p}
\spcend ,
\end{equation}
where $\delta_{ij}$ is Kronecker delta function.
In order to ensure the perfect reconstruction of a signal synthesised from its wavelet coefficients, one requires
the admissibility condition 
  \begin{equation}
  \label{eqn:admiss_full}
  0 <
  \admissc_\wav^\el \equiv
  \frac{8\pi^2}{2\el+1}
  \sum_{\m=-\el}^\el 
  \int_0^\infty
  \frac{\dx\scalea}{\scalea^3}
  \mid (\shcoeff{\wav_\scalea})_\elm \mid^2
  < \infty
  \end{equation}
to hold for all $\el \in \natno$. 
A proof of the admissibility criterion is given by \cite{wiaux:2005}.
Practically, it is
difficult to apply \eqn{\ref{eqn:admiss_full}} directly, thus a
necessary (and almost sufficient) condition for admissibility is the
zero-mean \mbox{condition \cite{antoine:1999}}
\begin{equation}
\admissc_\wav \equiv \int_\sphere  \dmu \: \frac{\wav(\sa)}{1+\cos\saa}=0
\spcend .
\end{equation}
In the Euclidean limit this condition naturally reduces to the
necessary zero-mean condition for Euclidean wavelets
\cite{wiaux:2005}.

\subsection{Correspondence principle and spherical wavelets}
\label{sec:wavelets}

The correspondence principle between spherical and Euclidean wavelets
states that the inverse stereographic projection of an
\emph{admissible} wavelet on the plane yields an 
\emph{admissible} wavelet on the unit sphere.  This result is proved by
\cite{wiaux:2005}.
%
Hence, mother spherical wavelets may be constructed from the projection
of mother Euclidean wavelets on the plane: 
\begin{equation}
\label{eqn:wav_proj}
\wav(\sa) = [\spo^{-1}\wav_{\realno^2}](\sa)
\spcend ,
\end{equation}
where $\wav_{\realno^2} \in L^2(\realno^2,\dx^2\xvect)$ is an admissible wavelet in the plane.
Directional spherical wavelets may be naturally constructed in this 
setting -- they are simply the projection of directional Euclidean planar
wavelets on to the sphere.

\begin{figure*}
\begin{minipage}{\textwidth}
\centering
\ifbw
  \mbox{
  \subfigure[Spherical Mexican hat wavelet (\smhw)]{\includegraphics[width=\mapwidth]{figures_bw/mexh_d0-2_moll_bw}} \quad
  \subfigure[Spherical butterfly wavelet (\sbw)]{\includegraphics[width=\mapwidth]{figures_bw/bfly_d0-2_moll_bw}} \quad
  \subfigure[Spherical real Morlet wavelet (\smw)]{\includegraphics[width=\mapwidth]{figures_bw/mrlt_d0-2_moll_bw}}
  }
\else
  \mbox{
  \subfigure[Spherical Mexican hat wavelet (\smhw)]{\includegraphics[width=\mapwidth]{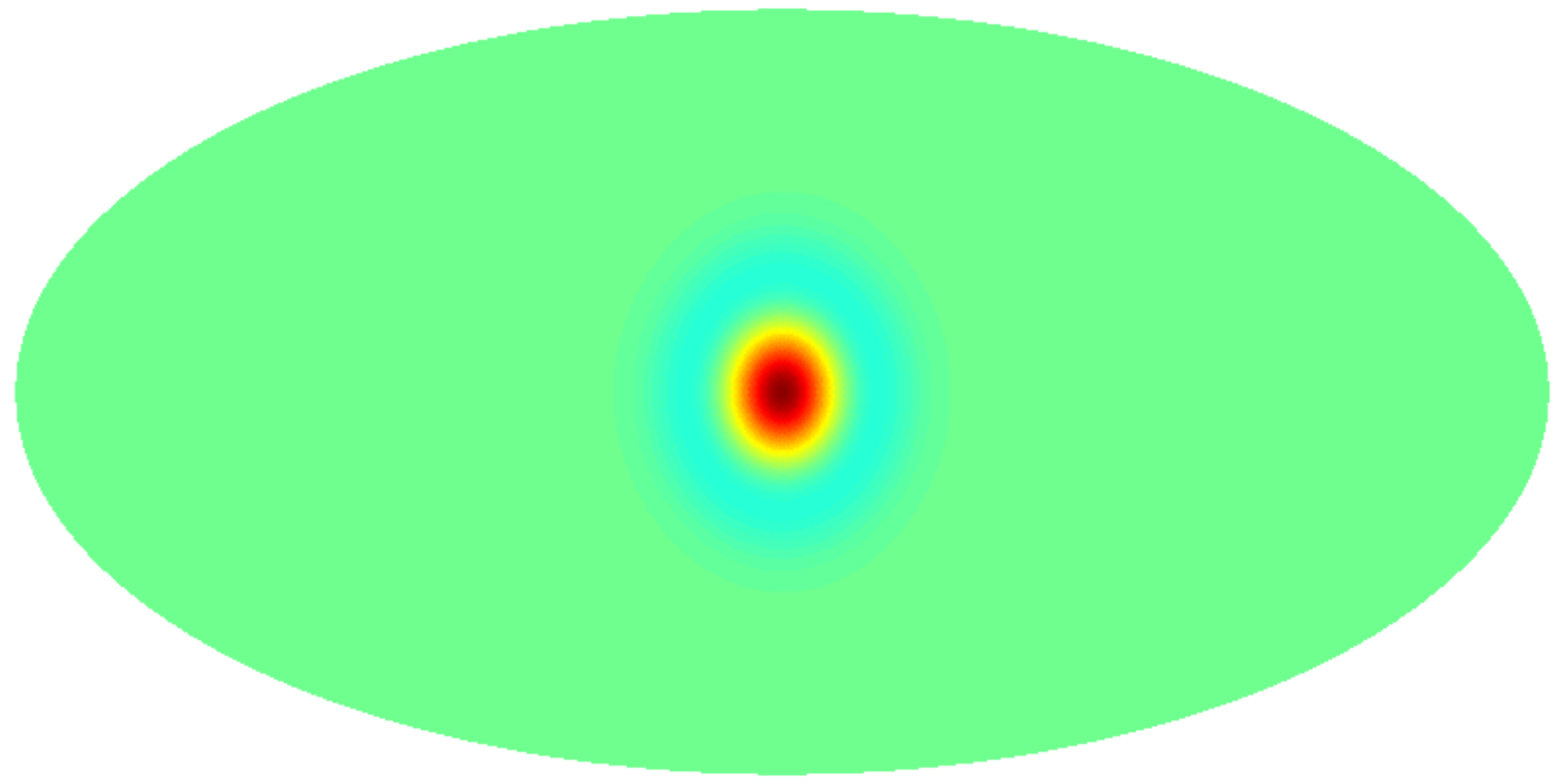}} \quad
  \subfigure[Spherical butterfly wavelet (\sbw)]{\includegraphics[width=\mapwidth]{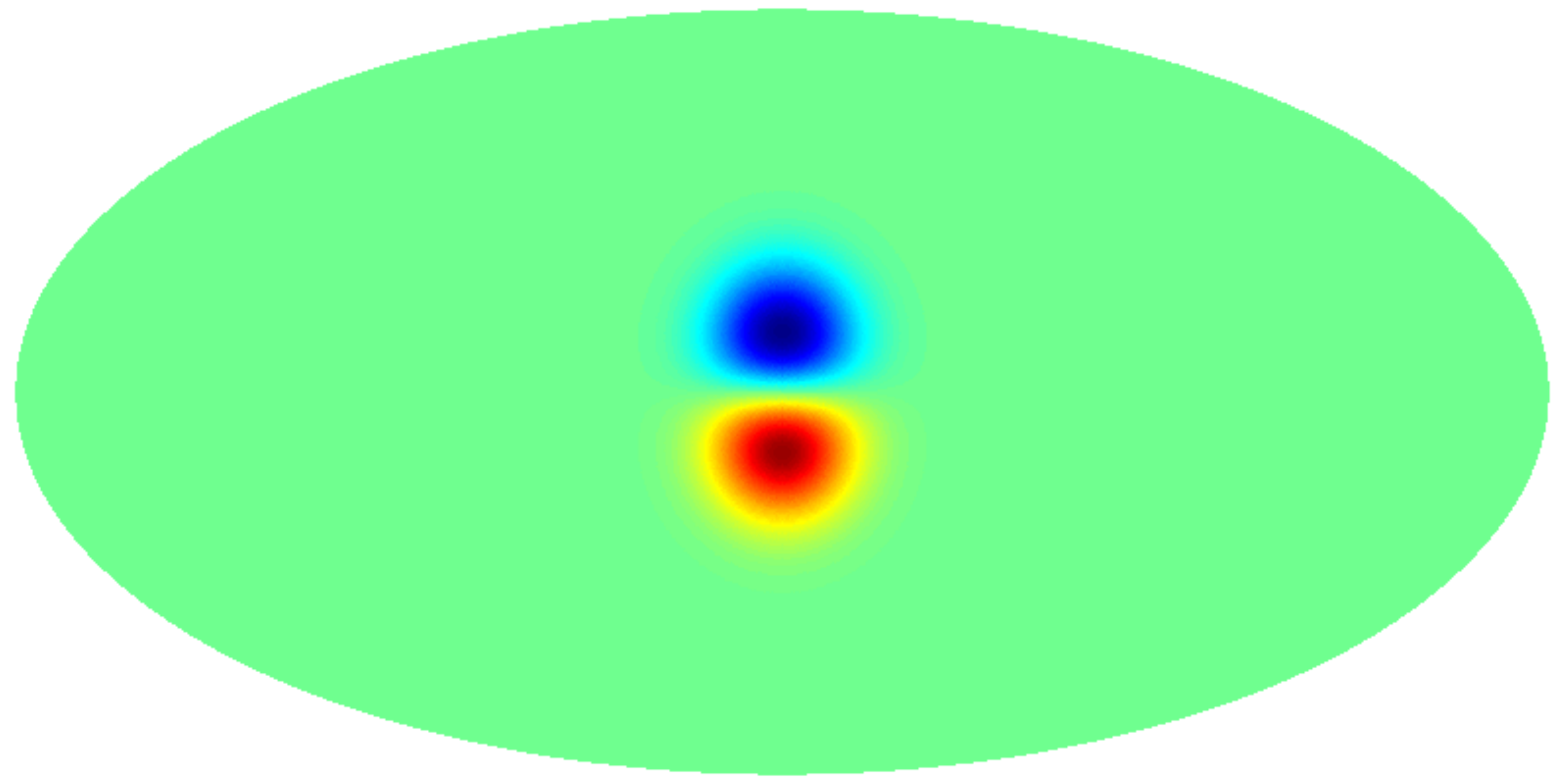}} \quad
  \subfigure[Spherical real Morlet wavelet (\smw)]{\includegraphics[width=\mapwidth]{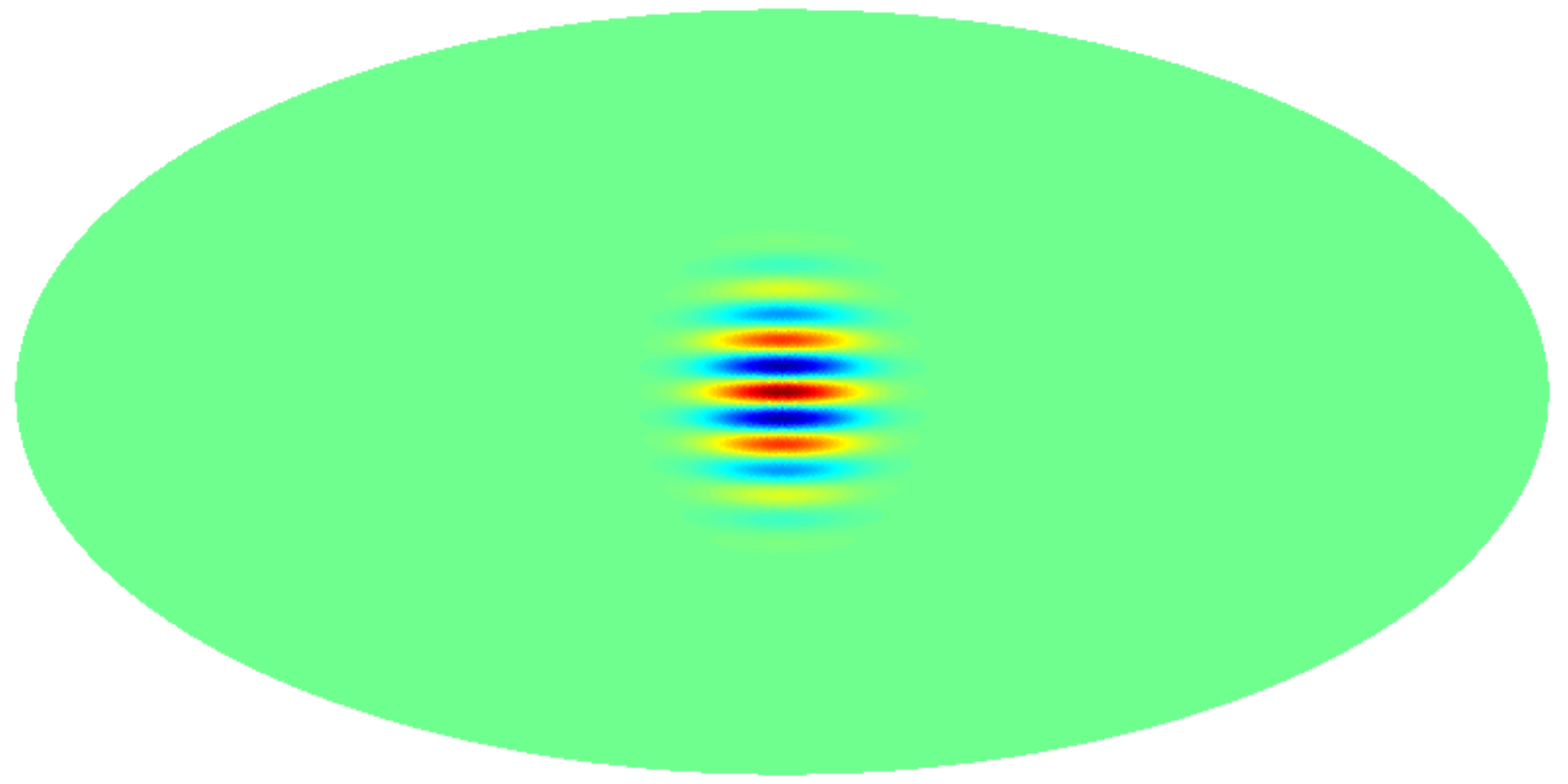}}
  }
\fi 
\caption{Spherical wavelets at scale $\scalea=\scaleb=0.2$.  Wavelet
  maps are displayed in the Mollweide projection, where the wavelets
  have been rotated down from the north pole for ease of observation.
  The \smw\ is plotted for wave vector $\kvect=(10,0)^T$.}
\label{fig:mother_wavelets}
\end{minipage}
\end{figure*}

We give examples of three spherical wavelets: the spherical Mexican
hat wavelet (\smhw); the spherical butterfly wavelet (\sbw); and
the spherical real Morlet wavelet (\smw).  These spherical wavelets are
illustrated in \fig{\ref{fig:mother_wavelets}}.
Each spherical wavelet is constructed by the stereographic projection
of the corresponding Euclidean wavelet onto the sphere, where the
Euclidean planar wavelets are defined by 
\begin{displaymath}
\wav_{\realno^2}^{\rm \smhw}(r,\sab) = 
\frac{1}{2}
(2 - r^2) \, e^{-r^2/2}
\spcend ,
\end{displaymath}
\begin{displaymath}
\wav_{\realno^2}^{\rm \sbw}(x,y) = 
x \, e^{-(x^2+y^2)/2}
\end{displaymath}
and
\begin{displaymath}
\wav_{\realno^2}^{\rm \smw}(\xvect; \kvect) = 
{\rm Re} \left(
e^{ \img \kvect \cdot \xvect / \sqrt{2}} \:
e^{ - \| \xvect \|^2 /2}
\right)
\end{displaymath}
respectively, where $\kvect$ is the wave vector of the \smw. 
The \smhw\ is proportional to the Laplacian of a Gaussian,
whereas the \sbw\ is proportional to the first partial derivative of a
Gaussian in the $x$-direction.  The \smw\ is a Gaussian modulated
sinusoid, or Gabor wavelet.

A full directional wavelet analysis on the unit sphere for large data sets
has previously been prohibited by the computational infeasibility of any
implementation. 
The computational burden of computing many orientations may be
reduced by using steerable wavelets, for which any continuous
orientation can be computed from a small number of basis
orientations \cite{wiaux:2005}.  This is achieved since steerable wavelets have a limited azimuthal band limit and may thus be represented as a finite sum of trigonometric exponentials \cite{wiaux:2005}.
However, in this case one must still
compute the initial transform for more than one orientation,
so although the computational burden is reduced, it is still
significant.  Moreover, we also require a fast approach for general 
non-steerable, directional wavelets.
We address this problem in the following section by
presenting a fast algorithm to perform the directional \cswt.

\section{Algorithms}
\label{sec:algorithms}

A range of algorithms of varying computational efficiency and numerical
accuracy are presented to perform the \cswt\ described in
\sectn{\ref{sec:cswt}}.  
We implement these algorithms in Fortran 90 and subsequently compare
computational complexity and typical execution time. 
The synthesis of a signal from its wavelet coefficients is not
considered any further.
Without loss of generality we consider only a single dilation
(\ie\ fixed \scalea\ and \scaleb).
 
\subsection{Tessellation schemes}
\label{sec:pixelisations}

It is necessary to discretise both the spherical coordinates of a
function defined on the unit sphere and also the Euler angle representation
of the \sothree\ rotation group.
%
The fast algorithms we present are performed in harmonic space and
hence are tessellation independent, provided an appropriate spherical
harmonic transform is defined for the tessellation.  However, the
semi-fast algorithm is restricted to an equi-angular
tessellation of the sphere.  
The various tessellation schemes adopted are defined below.

The equi-angular tessellation (also known as the equidistant
cylindrical projection (ECP)) of the spherical coordinates is defined
by 
$
\gridecp = \{ \saa_\ith=\frac{\pi\ith}{\nth}, \,
\sab_\ith=\frac{2\pi\iph}{\nph} :
0 \leq \ith \leq \nth - 1, \,
0 \leq \iph \leq \nph - 1
 \}
$. 
Let $\npix=\nth\nph$ denote the number of pixels in the tessellation.

We also consider the \healpix\ tessellation scheme since it is
commonly used for astrophysical data-sets of the CMB.  Pixels in the
\healpix\ scheme are of equal area and are located on rings of
constant latitude (the latter feature enables fast spherical harmonic 
transforms on the pixelised grid).  We refer the reader to
\cite{gorski:2005} for details of the \healpix\ scheme and here just
define the \healpix\ grid in terms of pixel indices:
$
\gridhpix = \{ 
(\sas)_\ipix = (\saa_\ipix, \sab_\ipix) :
0 \leq \ipix \leq \npix-1
\}
$.
The \healpix\ resolution is parameterised by \nside, where
$\npix=12\nside^2$.
It should be noted that an exact quadrature formula does not exist for the \healpix\ tessellation, thus spherical harmonic transforms are necessarily approximate.  This is not the case for the ECP or other practical tessellations (\eg\ IGLOO and GLESP) where exact quadrature may be performed.

The Euler angle domain of the spherical wavelet coefficients is in
general arbitrary, however we use the equi-angular discretisation
defined by
$
\grideula = \{ 
\eula_\ia=\frac{2\pi\ia}{\na}, \,
\eulb_\ib=\frac{\pi\ib}{\nb}, \,
\eulc_\ig=\frac{2\pi\ig}{\ngm} :
0 \leq \ia \leq \na - 1, \,
0 \leq \ib \leq \nb - 1, \,
0 \leq \ig \leq \ngm - 1 
\}
$.
Our fast algorithm, however, requires (for convenience) the
tessellation 
$
\grideulb = \{ 
\eula_\ia=\frac{2\pi\ia}{\na}, \,
\eulb_\ib=\frac{2\pi\ib}{\nb}, \,
\eulc_\ig=\frac{2\pi\ig}{\ngm} :
0 \leq \ia \leq \na - 1,  \,
0 \leq \ib \leq 2\nb - 1, \,
0 \leq \ig \leq \ngm - 1 
\}
$,
where the \eulb\ sampling is repeated.
Evaluating \eulb\ over the range $0$ to $2\pi$ is redundant, covering
the \sothree\ manifold exactly twice.  Nonetheless, the use of our fast
algorithm requires this range.  
Such an approach is not uncommon, as \cite{driscoll:1994,healy:2003}
also oversample a function of the sphere in the \saa\ direction in
order to develop fast spherical harmonic transforms on equi-angular grids. 
The overhead associated with our inefficient discretisation is more than
offset by the fast algorithm it affords, as described in
\sectn{\ref{sec:alg_fast}}.

\subsection{Direct case}

The \cswt\ defined by \eqn{\ref{eqn:cswt}} may be implemented directly
by applying an appropriate quadrature rule.
Using index subscripts to denote sampled signals, the direct \cswt\
implementation is given by 
\ifcol
  \begin{eqnarray}
  \label{eqn:alg_direct}
  \lefteqn{(\skywav)_{\ia,\ib,\ig} =}  \nonumber \\
  &&\displaystyle \sum_{\ipix=0}^{\npix-1}
  \left[\rot\left(\frac{2\pi\ia}{\na},\frac{\pi\ib}{\nb},\frac{2\pi\ig}{\ngm}\right)%
  \wav\right]_{\ipix}^\cconj
  \sky_{\ipix} \:
  \weight_\ipix
  \spcend ,
  \end{eqnarray}
\else
  \begin{equation}
  \label{eqn:alg_direct}
  (\skywav)_{\ia,\ib,\ig} = \sum_{\ipix=0}^{\npix-1}
  \left[\rot\left(\frac{2\pi\ia}{\na},\frac{\pi\ib}{\nb},\frac{2\pi\ig}{\ngm}\right)%
  \wav\right]_{\ipix}^\cconj 
  \sky_{\ipix} \:
  \weight_\ipix
  \spcend ,
  \end{equation}
\fi
where the pixel sum is over all the pixels of any chosen tessellation.
The weights for the ECP \gridecp\ grid are given by 
$\weight_\ipix=\weight_{\ith}=\frac{2\pi^2 \sin{\saa_\ith}}{\nth\nph}$,
whereas the equal pixel areas of the \healpix\ \gridhpix\ grid ensure
the pixel weights, given by $\weight_\ipix=\frac{4\pi}{\npix}$, are
independent of position.  

Discretisation techniques other than the plain Riemann sum used in
\eqn{\ref{eqn:alg_direct}} would be beneficial only if additional
regularity conditions are imposed on the signal \sky\
\cite{antoine:2002}.  It is also possible to choose other weights to
achieve a better approximation of the integral.  An example of a
different equi-angular discretisation and a different choice for the
weights is given by the sampling theorem for band-limited functions on
the sphere developed by \cite{healy:2003}. 

Evaluation of \eqn{\ref{eqn:alg_direct}} requires the computation of a
\mbox{2-dimensional} summation, evaluated over a \mbox{3-dimensional} grid.  We
assume the number of samples for each discretised angle, except
$\gamma$, is of the same order \n.  Typically the number of samples
in the $\gamma$ direction is of a much lower order, so we treat this
term separately.  The complexity of the direct algorithm is 
$\order(\ngm\n^4)$.


\subsection{Semi-fast case}

We rederive here the semi-fast implementation of the \cswt\ described
by \cite{antoine:2002} and implemented using the \yawtb\ Matlab wavelet toolbox and SpharmonicKit.
This algorithm involves performing a separation of variables so that
one rotation may be performed in Fourier space.  The algorithm is
restricted to the equi-angular grid \gridecp\ (in essence pixels must
only be defined on equal latitude rings, however some form of
interpolation and down-sampling is then required to extract samples
for equal longitudes). 

Firstly, the \eula\ rotation is represented by shifting the corresponding
wavelet samples to give 
\ifcol
  \begin{eqnarray*}
  \lefteqn{(\skywav)_{\ia,\ib,\ig} =} \nonumber \\
  &&
  \displaystyle
  \sum_{\ith=0}^{\nth-1}
  \sum_{\iph=0}^{\nph-1}
  \left[\rot\left(0,\frac{\pi\ib}{\nb},\frac{2\pi\ig}{\ngm}\right)
  \wav\right]_{\ith,\iph-\ia}^\cconj \!\!
  \sky_{\ith,\iph} 
  \weight_\ith
  \spcend ,
  \end{eqnarray*}
\else
  \begin{displaymath}
  (\skywav)_{\ia,\ib,\ig} = 
  \sum_{\ith=0}^{\nth-1}
  \sum_{\iph=0}^{\nph-1}
  \left[\rot\left(0,\frac{\pi\ib}{\nb},\frac{2\pi\ig}{\ngm}\right)
  \wav\right]_{\ith,\iph-\ia}^\cconj \:
  \sky_{\ith,\iph} \:
  \weight_\ith
  \spcend ,
  \end{displaymath}
\fi
where the index \iph\ is extended periodically with period \nph.  The
discrete space convolution theorem may then be applied to represent
the inner summation as the inverse discrete Fourier transform (\dft) of
the product of the wavelet and signal \dft\ samples (note that only a
\mbox{1-dimensional} \dft\ is performed in the azimuthal direction):
\ifcol
  \begin{eqnarray}
  \label{eqn:alg_semi}
  \lefteqn{(\skywav)_{\ia,\ib,\ig} =} \nonumber \\ 
  &&
  \sum_{\ith=0}^{\nth-1}
  \left\{
  \frac{1}{\nph} \right.
  \sum_{\ik=0}^{\nph-1}
  \mathcal{F}^\cconj
  \left[\rot\left(0,\frac{\pi\ib}{\nb},\frac{2\pi\ig}{\ngm}\right)
  \wav\right]_{\ith,\ik} \nonumber \\
  &&\left. \times \:
  {\mathcal{F}(\sky)}_{\ith,\ik} \:
  e^\frac{\img2\pi\ik\iph}{\nph}
  \right\}
  \weight_\ith
  \spcend ,
  \end{eqnarray}
\else
  \begin{equation}
  \label{eqn:alg_semi}
  (\skywav)_{\ia,\ib,\ig} = 
  \sum_{\ith=0}^{\nth-1}
  \left\{
  \frac{1}{\nph}
  \sum_{\ik=0}^{\nph-1}
  \mathcal{F}^\cconj
  \left[\rot\left(0,\frac{\pi\ib}{\nb},\frac{2\pi\ig}{\ngm}\right)
  \wav\right]_{\ith,\ik} \:
  {\mathcal{F}(\sky)}_{\ith,\ik} \:
  e^\frac{\img2\pi\ik\iph}{\nph}
  \right\}
  \weight_\ith
  \spcend ,
  \end{equation}
\fi
where $\mathcal{F}(\cdot)_{n,k}$ denotes \mbox{1-dimensional} \dft\ coefficients.
A fast Fourier transform (\fft) may then be applied to evaluate
simultaneously all of the $\ia$ terms of the expression enclosed in the 
curly braces in \eqn{\ref{eqn:alg_semi}}.  A final
summation (integral) over $\ith$ produces the spherical wavelet
coefficients for a given $\ib$ and $\ig$, for \emph{all} $\ia$.
Applying an \fft\ to evaluate simultaneously one summation rapidly,
reduces the complexity of the \cswt\ implementation to 
$\order(\ngm \n^3 \log_2 \n  )$.



\subsection{Fast azimuthally symmetric case}

The fast azimuthally symmetric \cswt\ algorithm is posed in harmonic space, where 
$\shcoeff{s}_\elm = \left< \sh_\elm \mid s \right>$ are the spherical harmonic
coefficients of a function $s \in L^2(\sphere,\dmu)$, as described in
\sectn{\ref{sec:wavelet_transform}}.  For the special case where the
wavelet is azimuthally symmetric (\ie\ invariant under azimuthal rotations), it is essentially only a function of \saa\ and may be represented in terms of its Legendre expansion.
In this case the 
harmonic representation of the wavelet coefficients is given by
the product of the signal and wavelet spherical harmonic coefficients:
\begin{equation}
\label{eqn:fast_isotropic}
(\shcoeff{\skywav})_\elm = 
\sqrt{\frac{4 \pi}{2\el+1}} \:
\shcoeff{\wav}_{\el 0}^\cconj \:
\shcoeff{\sky}_\elm
\spcend ,
\end{equation}
noting that the harmonic coefficients of an azimuthally symmetric wavelet are zero
for $\m\neq0$.  In practice, one requires that at least one of the
signals, usually the wavelet, has a finite band limit so that
negligible power is present in those coefficients above a certain
\elmax.  We then only need to consider $\el \leq \elmax$ (a detailed discussion of the determination of \elmax\ is presented in \cite{bogdanova:2004}).
Once the spherical harmonic representation of the wavelet coefficients
is calculated by \eqn{\ref{eqn:fast_isotropic}}, the inverse spherical
harmonic transform is applied to compute the wavelet coefficients in
the Euler domain. 
%
The complexity of the fast isotropic \cswt\ algorithm is dominated by
the spherical harmonic transforms.  
For a tessellation containing pixels on rings of constant latitude, a fast spherical harmonic transform may be performed (see \eg \cite{mohlenkamp:1997,mohlenkamp:1999,gorski:2005}).  This reduces the complexity of the spherical harmonic transform from $\order(\n^4)$ to $\order(\n^3)=\order(\npix^{3/2})$. 
For certain tessellation schemes fast spherical harmonic transforms of lower complexity are also available, however these are related directly to the tessellation (\eg\ \cite{mohlenkamp:1997,mohlenkamp:1999,driscoll:1994,healy:2003}).  In particular, the algorithm developed by \cite{driscoll:1994,healy:2003} for the ECP tessellation scales as $\order(\n^2\log\n)$.

The fast azimuthally symmetric \cswt\ algorithm is posed purely in harmonic space and consequently 
the algorithm is tessellation independent.  However, we are
restricted to azimuthally symmetric wavelets and lose the ability to
perform the directional analysis inherent in the wavelet transform
construction.  

\subsection{Fast directional case}
\label{sec:alg_fast}

We present the most general fast directional \cswt\ algorithm for
non-azimuthally symmetric wavelets, \ie\ steerable and directional wavelets, in this section.
Again, the algorithm is posed purely in harmonic space and so 
is tessellation independent.  We do, however, use the
equi-angular \grideulb\ discretisation of the wavelet coefficient
domain, although other discretisations may be used if \fft s are also 
defined on these grids.
The \cswt\ at a particular scale (\ie\ a particular \scalea\ and
\scaleb) is essentially a spherical 
{con\-volu\-tion}, hence we apply the fast spherical
convolution algorithm proposed by \cite{wandelt:2001} to evaluate
the wavelet transform.   
The algorithm proceeds by factoring the rotation into two separate
rotations, each of which involves only a constant polar rotation component. 
Azimuthal rotations may then be performed in harmonic space at far
less computation expense than polar rotations. 
We subsequently rederive the fast spherical convolution algorithm
developed by \cite{wandelt:2001}, as applied to our application of
evaluating the \cswt.
The harmonic representation of the \cswt\ is first presented, followed
by the discretisation and fast implementation.

\subsubsection{Harmonic formulation}

Substituting the spherical harmonic expansions of the wavelet and signal
into the wavelet transform defined by \eqn{\ref{eqn:cswt}}, and noting the
orthogonality of the spherical harmonics described by \eqn{\ref{eqn:shortho}},
yields the harmonic representation
\ifcol
  \begin{eqnarray}
  \label{eqn:harmonic:1}
  \lefteqn{\skywav(\eula, \eulb, \eulc) =} \nonumber \\
  &&
  \sum_{\el =0}^{\elmax} \:
  \sum_{\m=-\el}^{\el} \:
  \sum_{\m\p=-\el}^{\el}
  \left [
  {\dmatbig_{\m\m\p}^{\el }(\euls) \,
  \shcoeff{\wav}_{\el \m\p}} \right ]^\cconj 
  \shcoeff{\sky}_{\el \m}
  \spcend .
  \end{eqnarray}
\else
  \begin{equation}
  \label{eqn:harmonic:1}
  \skywav(\eula, \eulb, \eulc) =
  \sum_{\el =0}^{\elmax} \:
  \sum_{\m=-\el}^{\el} \:
  \sum_{\m\p=-\el}^{\el}
  \left [
  {\dmatbig_{\m\m\p}^{\el }(\euls) \,
  \shcoeff{\wav}_{\el \m\p}} \right ]^\cconj \:
  \shcoeff{\sky}_{\el \m}
  \spcend .
  \end{equation}
\fi
Again, we assume negligible power above \elmax\ in at least one of the
signals, usually the wavelet, so that the outer summation is truncated
to \elmax. 
The additional summation over $\m\p$ and the $\dmatbig_{mm\p}^\el$
Wigner rotation matrices that are introduced
characterise the rotation of a spherical harmonic, noting that a rotated
spherical harmonic may be represented by a sum of 
harmonics of the same $\el$ \cite{brink:1993,ritchie:1999}:
\begin{displaymath}
\left[\rot(\euls)\sh_{\elm}\right](\sa) = 
\sum_{\m\p=-\el}^{\el} 
\dmatbig_{\m\p\m}^{\el}(\euls) \: \sh_{\el \m\p}(\sa)
\spcend .
\end{displaymath}
The Wigner rotation matrices may be decomposed as \cite{brink:1993,ritchie:1999}
\begin{equation}
\label{eqn:d_decomp}
\dmatbig_{\m\m\p}^{\el}(\euls)
= e^{-\img \m\eula} \:
\dmatsmall_{\m\m\p}^\el(\eulb) \:
e^{-\img \m\p\eulc}
\spcend ,
\end{equation}
where the real polar \dmatsmall-matrix is defined by \cite{brink:1993}
\ifcol
  \begin{eqnarray*}
  \lefteqn{\dmatsmall^\el_{\m \m\p}(\eulb)  = 
  \sum_{t = \max(0, \m - \m\p)}^{\min(\el + \m, \el - \m\p)} (-1)^t} \nonumber\\
  && \times \frac{
  \left[(\el+\m)! \, (\el-\m)! \, (\el+\m\p)! \, (\el-\m\p)! \, \right]^{1/2}}
  {(\el+\m-t)! \, (\el-\m\p-t)! \, (t+\m\p-\m)! \, t!} \nonumber \\
  && \times \left[\cos\!\left(\frac{\eulb}{2} \right)\right]^{2\el+\m-\m\p-2t}
  \left[\sin \! \left(\frac{\eulb}{2} \right)\right]^{\m\p-\m+2t}
  \spcend ,
  \end{eqnarray*}
\else
  \begin{eqnarray*}
  \dmatsmall^\el_{\m \m\p}(\eulb) & = & 
  \sum_{t = \max(0, \m - \m\p)}^{\min(\el + \m, \el - \m\p)} (-1)^t
  \frac{
  \left[(\el+\m)! \, (\el-\m)! \, (\el+\m\p)! \, (\el-\m\p)! \, \right]^{1/2}}
  {(\el+\m-t)! \, (\el-\m\p-t)! \, (t+\m\p-\m)! \, t!} \nonumber \\
  && \times \left[\cos\!\left(\frac{\eulb}{2} \right)\right]^{2\el+\m-\m\p-2t}
  \left[\sin \! \left(\frac{\eulb}{2} \right)\right]^{\m\p-\m+2t}
  \spcend ,
  \end{eqnarray*}
\fi
and the sum over $t$ is defined so that the arguments of the factorials
are non-negative.  Recursion formulae are available to compute rapidly the Wigner rotation matrices in the basis of either complex \cite{risbo:1996,choi:1999} or real \cite{ivanic:1996,blanco:1997} spherical harmonics.  We employ the recursion formulae described in \cite{risbo:1996} in our implementation.
The decomposition shown in \eqn{\ref{eqn:d_decomp}} is exploited by
factoring the rotation $\rot(\euls)$ into two separate rotations,
both of which only contain 
a constant $\pm \pi/2$ polar rotation:
\ifcol
  \begin{eqnarray}
  \label{eqn:rot_factor}
  \rot(\euls)
  &=& \rot(\eula-\pi/2, \; -\pi/2, \; \eulb) \nonumber \\
  &&\times \:
  \rot(0, \; \pi/2, \; \eulc+\pi/2)
  \spcend .
  \end{eqnarray}
\else
  \begin{equation}
  \label{eqn:rot_factor}
  \rot(\euls)
  = \rot(\eula-\pi/2, \; -\pi/2, \; \eulb) \:\:
  \rot(0, \; \pi/2, \; \eulc+\pi/2)
  \spcend .
  \end{equation}
\fi
By factoring the rotation in this manner and applying the decomposition
described by \eqn{\ref{eqn:d_decomp}}, \eqn{\ref{eqn:harmonic:1}} 
can be rewritten as
\ifcol
  \begin{eqnarray}
  \label{eqn:harmonic:2}
  \lefteqn{\skywav(\euls) =} \nonumber \\
  &&
  \sum_{\el=0}^{\elmax} \:
  \sum_{\m=-\el}^{\el} \:
  \sum_{\m\p=-\el}^{\el}
  \sum_{\m\pp=-\min(\mmax,\el)}^{\min(\mmax,\el)}
  \dmatsmall_{\m\p \m}^\el(\pi/2) \:
  \nonumber \\
  && \times \:
  \dmatsmall_{\m\p \m\pp}^\el(\pi/2) \:
  \shcoeff{\wav}_{\el \m\pp}^\cconj \:
  \shcoeff{\sky}_{\el \m} \nonumber \\
  && \times \:
  e^{\img [\m(\eula-\pi/2) + \m\p\eulb + 
  \m\pp(\eulc+\pi/2) ]}
  \spcend ,
  \end{eqnarray}
\else
  \begin{eqnarray}
  \label{eqn:harmonic:2}
  \skywav(\euls) &=&
  \sum_{\el=0}^{\elmax} \:
  \sum_{\m=-\el}^{\el} \:
  \sum_{\m\p=-\el}^{\el} \:
  \sum_{\m\pp=-\min(\mmax,\el)}^{\min(\mmax,\el)}
  \dmatsmall_{\m\p \m}^\el(\pi/2) \:
  \dmatsmall_{\m\p \m\pp}^\el(\pi/2)
  \nonumber \\
  & & 
  \times \:\:
  \shcoeff{\wav}_{\el \m\pp}^\cconj \:
  \shcoeff{\sky}_{\el \m} \:
  e^{\img [\m(\eula-\pi/2) + \m\p\eulb + 
  \m\pp(\eulc+\pi/2) ]}
  \spcend ,
  \end{eqnarray}
\fi
where the symmetry relationship
\mbox{$\dmatsmall_{\m \m\p}^{\el}(-\eulb)=\dmatsmall_{\m\p \m}^{\el}(\eulb)$}
has been applied.  
(A similar factoring of the rotation and harmonic space representation has been independently performed in \cite{bulow:2001}.)
Steerable wavelets may have a low effective band limit 
$\mmax \ll \elmax$, in which case the the inner summation in \eqn{\ref{eqn:harmonic:2}} may be truncated to \mmax.
For general directional wavelets this is not the case and one must use $\mmax = \elmax$.

Evaluating the harmonic formulation given by
\eqn{\ref{eqn:harmonic:2}} directly would be no more efficient that 
approximating the initial integral \eqn{\ref{eqn:cswt}} using
simple quadrature.  
However, 
\eqn{\ref{eqn:harmonic:2}}
is represented in such a way that the presence of complex exponentials
may be exploited such that \fft s may be applied to evaluate rapidly
the three summations simultaneously. 


\subsubsection{Fast implementation}

Azimuthal rotations may be applied with far less computational expense
than polar rotations since they appear within complex exponentials
in \eqn{\ref{eqn:harmonic:2}}.  Although the \dmatsmall-matrices can be
evaluated reasonably quickly and reliably using recursion formulae, the
basis for the fast implementation is to avoid these polar rotations as
much as possible and use \fft s to evaluate rapidly all of the azimuthal
rotations simultaneously. 
This is the motivation for factoring the rotation by
\eqn{\ref{eqn:rot_factor}} so that all Euler angles occur as azimuthal
rotations.

The discretisation of each Euler angle may in general be {arbi\-trary}.
However, to exploit standard \fft\ routines uniform sampling is
adopted, \ie\ grid \grideulb\ is used
(see \sectn{\ref{sec:pixelisations}}).  As mentioned, this
discretisation is inefficient since it covers the \sothree\ manifold
exactly twice, nevertheless it enables the use of standard \fft\
routines which significantly increases the speed of the algorithm. 
Discretising \eqn{\ref{eqn:harmonic:2}} in this manner and
interchanging the order of summation we obtain
\ifcol
  \begin{eqnarray*}
  \lefteqn{(\skywav)_{\ia, \ib, \ig}  =} \nonumber \\
  && \displaystyle 
  \sum_{\m=-\elmax}^{\elmax} \:
  \sum_{\m\p=-\elmax}^{\elmax} \:
  \sum_{\m\pp=-\mmax}^{\mmax} 
  \nonumber \\
  &&
  \displaystyle
  \sum_{\el=\max(\mid \m \mid, \mid \m\p \mid, \mid \m\pp \mid)}^{\elmax}
  \dmatsmall_{\m\p \m}^\el(\pi/2) \:
  \dmatsmall_{\m\p \m\pp}^\el(\pi/2) \:
  \shcoeff{\wav}_{\elm\pp}^\cconj \:
  \shcoeff{\sky}_{\elm} \nonumber \\
  \nonumber \\
  &&\times \:
  e^{\img [\m(2\pi\ia/\na - \pi/2) 
  + \m\p 2\pi\ib/\nb
  + \m\pp (2\pi\ig/\ngm + \pi/2) ]}
  \spcend .
  \end{eqnarray*}
\else
  \begin{eqnarray*}
  (\skywav)_{\ia, \ib, \ig} & = &
  \sum_{\m=-\elmax}^{\elmax} \:
  \sum_{\m\p=-\elmax}^{\elmax} \:
  \sum_{\m\pp=-\mmax}^{\mmax} 
  \sum_{\el=\max(\mid \m \mid, \mid \m\p \mid, \mid \m\pp \mid)}^{\elmax}
  \dmatsmall_{\m\p \m}^\el(\pi/2) \:
  \dmatsmall_{\m\p \m\pp}^\el(\pi/2)
  \nonumber \\
  & & 
  \times
  \shcoeff{\wav}_{\elm\pp}^\cconj \:
  \shcoeff{\sky}_{\elm} \:
  e^{\img [\m(2\pi\ia/\na - \pi/2) 
  + \m\p 2\pi\ib/\nb
  + \m\pp (2\pi\ig/\ngm + \pi/2) ]}
  \spcend .
  \end{eqnarray*}
\fi
Shifting the indices yields
\ifcol
  \begin{eqnarray}
  \label{eqn:cswt_fast}
  \lefteqn{(\skywav)_{\ia, \ib, \ig} =} \nonumber \\
  &&\displaystyle \sum_{\m=0}^{2\elmax} \:
  \sum_{\m\p=0}^{2\elmax} \:
  \sum_{\m\pp=0}^{2\mmax}
  e^{\img 2 \pi
  ( \ia\m/\na + \ib\m\p/\nb + \ig\m\pp/\ngm )} \nonumber \\
  &&\times \:
  \cswtfftterm_{\m,\m\p,\m\pp}
  \spcend ,
  \end{eqnarray}
\else
  \begin{equation}
  \label{eqn:cswt_fast}
  (\skywav)_{\ia, \ib, \ig} =
  \sum_{\m=0}^{2\elmax} \:
  \sum_{\m\p=0}^{2\elmax} \:
  \sum_{\m\pp=0}^{2\mmax}
  e^{\img 2 \pi
  ( \ia\m/\na + \ib\m\p/\nb + \ig\m\pp/\ngm )}
  \cswtfftterm_{\m,\m\p,\m\pp}
  \spcend ,
  \end{equation}
\fi
where
\ifcol
  \begin{eqnarray}
  \lefteqn{\cswtfftterm_{\m,\m\p,\m\pp} =
  e^{\img(\m\pp-\m)\pi/2}} \nonumber \\
  &&\times
  \displaystyle
  \sum_{\el=\max(\mid \m \mid, \mid \m\p \mid, \mid \m\pp \mid)}^{\elmax}
  \dmatsmall_{\m\p \m}^\el(\pi/2) \:
  \dmatsmall_{\m\p \m\pp}^\el(\pi/2) \nonumber \\
  &&\times \:
  \shcoeff{\wav}_{\el \m\pp}^\cconj \:
  \shcoeff{\sky}_{\el \m}
  \label{eqn:cswt_fast_term}
  \end{eqnarray}
\else
  \begin{equation}
  \cswtfftterm_{\m,\m\p,\m\pp} =
  e^{\img(\m\pp-\m)\pi/2}
  \sum_{\el=\max(\mid \m \mid, \mid \m\p \mid, \mid \m\pp \mid)}^{\elmax}
  \dmatsmall_{\m\p \m}^\el(\pi/2) \:
  \dmatsmall_{\m\p \m\pp}^\el(\pi/2) \:
  \shcoeff{\wav}_{\el \m\pp}^\cconj \:
  \shcoeff{\sky}_{\el \m}
  \label{eqn:cswt_fast_term}
  \end{equation}
\fi
is extended periodically.  
Note that the phase shift introduced by shifting the indices of the
summation in \eqn{\ref{eqn:cswt_fast}}, shifts the
$\cswtfftterm_{\m,\m\p,\m\pp}$ indices back.  
Making the associations \mbox{$\na=2\elmax+1$}, $\nb=2\elmax+1$ and
$\ngm=2\mmax+1$, one notices that \eqn{\ref{eqn:cswt_fast}} is 
the unnormalised \mbox{3-dimensional} inverse \dft\ of
\eqn{\ref{eqn:cswt_fast_term}}. 
Nyquist sampling in $(\euls)$ is inherently ensured by the
associations made with \elmax\ and \mmax.

The \cswt\ may be performed rapidly in spherical harmonic
space by constructing the \cswtfftterm-matrix of
\eqn{\ref{eqn:cswt_fast_term}} from spherical harmonic coefficients and
precomputed \mbox{\dmatsmall-matrices},
followed by the application of an \fft\ to 
evaluate rapidly all three Euler angles of the discretised \cswt\
simultaneously.  The complexity of the algorithm is dominated by the 
computation of the \cswtfftterm-matrix.  This involves performing a
\mbox{1-dimensional} summation over a \mbox{3-dimensional} grid, hence the
algorithm is of order $\order(\ngm\n^3)$.

Additional benefits may be realised for real signals by exploiting the
resulting conjugate symmetry relationship.  
Memory and computational requirements may be reduced by a
further factor of two for real signals by noting the conjugate
symmetry relationship
$\cswtfftterm_{-m,-m\p,-m\pp}=\cswtfftterm_{m,m\p,m\pp}^\cconj$.
In our implementation we use the 
complex-to-real \fft\ routines of the FFTW\footnote{\url{http://www.fftw.org/}}
package, which are approximately twice as fast as the
equivalent complex-to-complex routines.

\subsection{Comparison}

We summarise the computational complexities of the various \cswt\
algorithms for a 
single pair of scales and single orientation in
\tbl{\ref{tbl:complexity}}.   
The complexity of each algorithm scales with the
number of dilations considered and, for those algorithms that
facilitate a directional analysis (\ie\ all but the fast azimuthally symmetric
algorithm), with the number of orientations considered.  
The most general fast directional algorithm provides a
saving of $\order(\n)$ over the direct case, where the number of
pixels on the sphere and the harmonic band limit are related to \n\ by 
$\order(\npix)=\order(\elmax^2)=\order(\n^2)$.

\begin{table}
  \caption{Algorithm complexity for one scale and one orientation.}
\label{tbl:complexity}
\begin{center}
\begin{tabular}{ll} \cline{1-2}
Algorithm & Complexity \\ \cline{1-2}
Direct           & $\order(\n^4)$ \\
Semi-fast        & $\order(\n^3\log_2\n)$ \\
Fast azimuthally symmetric   & $\order(\n^3)$ \\
Fast directional & $\order(\n^3)$ \\ \cline{1-2}
\end{tabular}
\end{center}
\end{table}

We implement all algorithms in Fortran 90, adopting the \healpix\ tessellation of the sphere
(which, incidentally, 
is the tessellation scheme of the WMAP CMB data \cite{bennett:2003}).
Typical execution times for the 
algorithms are tabulated in \tbl{\ref{tbl:exec_time}} for a range of 
resolutions from 110 down to 3.4 arcminutes.
The improvements provided by the fast
algorithms are apparent.  Indeed, it is not feasible to run the direct
algorithm on data-sets with a resolution much greater than
$\npix \simeq 5\times 10^5$.
For data-sets of practical size, such as the WMAP 
($\npix \simeq 3\times10^6$) and
forthcoming Planck ($\npix \simeq 50\times10^6$) CMB data, the fast
implementations of the \cswt\ are essential.

The semi-fast algorithm is also implemented using the \healpix\
tessellation.  However, to perform the outer summation (integration)
continual interpolation followed by down-sampling is required on each
iso-latitude ring to essentially resample the data on an ECP tessellation. 
This increases the execution 
time of the implementation of the semi-fast algorithm on the \healpix\
grid to an extent that the semi-fast algorithm provides little
advantage over the direct algorithm.  To appreciate the advantages of
the semi-fast approach it must be implemented on an ECP tessellation,
hence we do not tabulate the execution times for our implementation of
this algorithm on the \healpix\ grid as it provides an unfair
comparison. 

It is also important to note that although complexity scales with the
number of dilations and orientations considered, 
execution time does not for the fast algorithms.
Execution time does scale in this manner for
the direct algorithm.
There are a number of additional overheads
associated with the fast algorithms, such as computing spherical
harmonic coefficients and \dmatsmall-matrices.  Consequently, the fast
algorithms provide additional execution time savings that are not
directly apparent in \tbl{\ref{tbl:exec_time}}.  For example, the
execution time of the fast azimuthally symmetric and directional algorithms for 10
dilations at a resolution of $\npix \simeq 8 \times 10^5$
($\nside=256$) are 3:08.20 
and 7:06.83 (minutes:seconds) respectively, which is considerably
faster than ten times the execution time of one dilation.

\begin{table}
  \caption{Typical execution time (minutes:seconds) for each algorithm
    run on an Intel P4-M 3GHz laptop with 512MB of memory.}
\label{tbl:exec_time}
\begin{center}
\begin{tabular}{rrrrr} \cline{1-5}
\multicolumn{2}{c}{Resolution} & \multicolumn{3}{c}{Algorithm
  execution time} \\
\multicolumn{1}{c}{\nside} & \multicolumn{1}{c}{\npix} & Direct & Fast isotropic & Fast directional \\ \cline{1-5}
  32 &     12,288 & 3:25.37         & 0:00.06  &  0:00.10 \\
  64 &     49,152 & 54:31.75        & 0:00.38  &  0:00.74 \\
 256 &    786,432 & -- \hspace{3mm} & 0:28.00  &  0:52.55 \\
 512 &  3,257,292 & -- \hspace{3mm} & 3:43.69  &  7:57.75 \\
1024 & 12,582,912 & -- \hspace{3mm} & 28:23.85 & 71:31.68 \\\cline{1-5}
\end{tabular}
\end{center}
\end{table}


\section{Application}
\label{sec:application}

\begin{figure}
\centering
\ifbw
\else
  \subfigure[Binary Earth map]{\includegraphics[width=\coeffwidth]{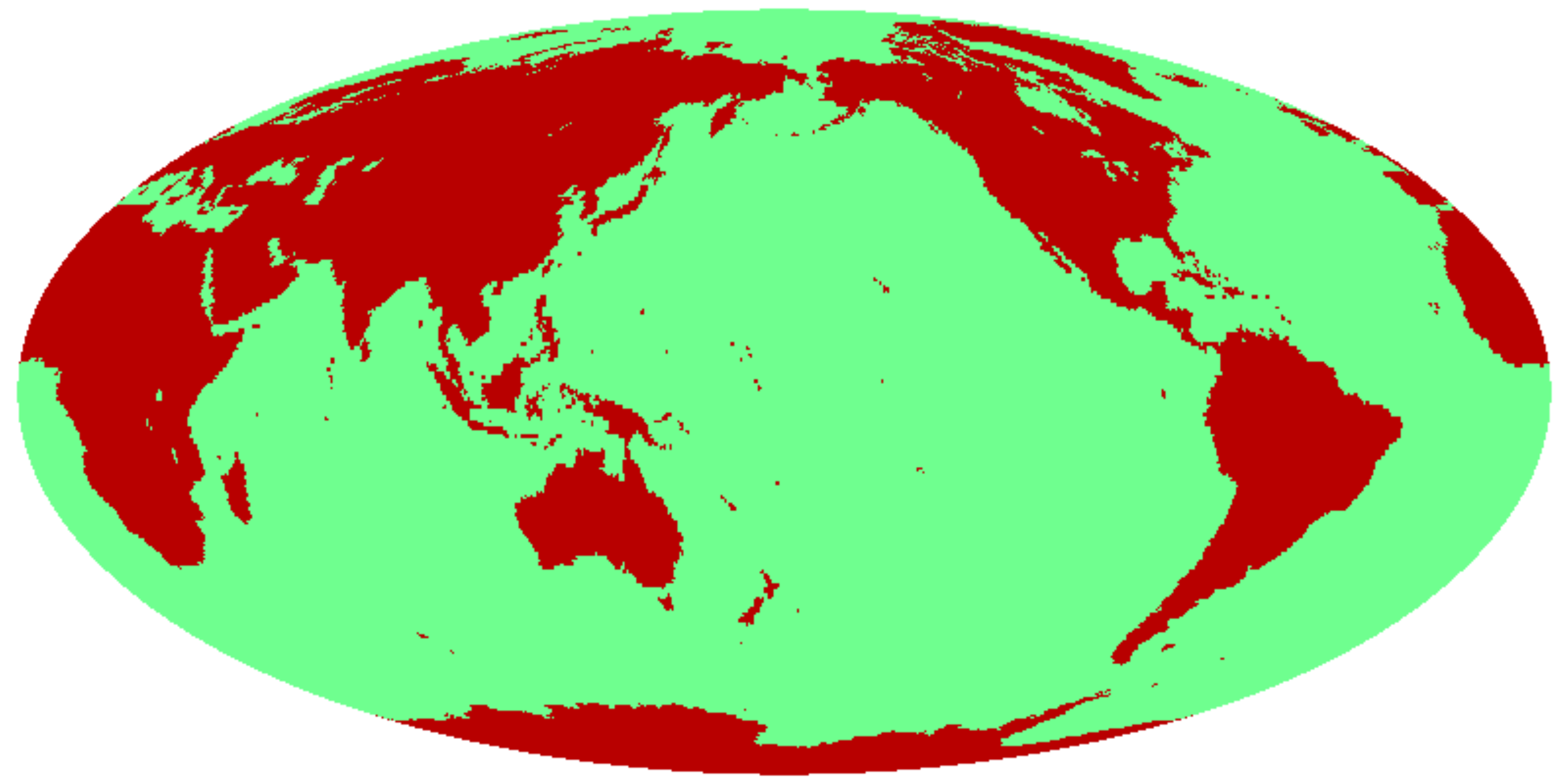}}
  \subfigure[Spherical butterfly wavelet coefficients for scale $\scalea=\scaleb=0.03$ and orientation $\eulc=0^\circ$]{\includegraphics[width=\coeffwidth]{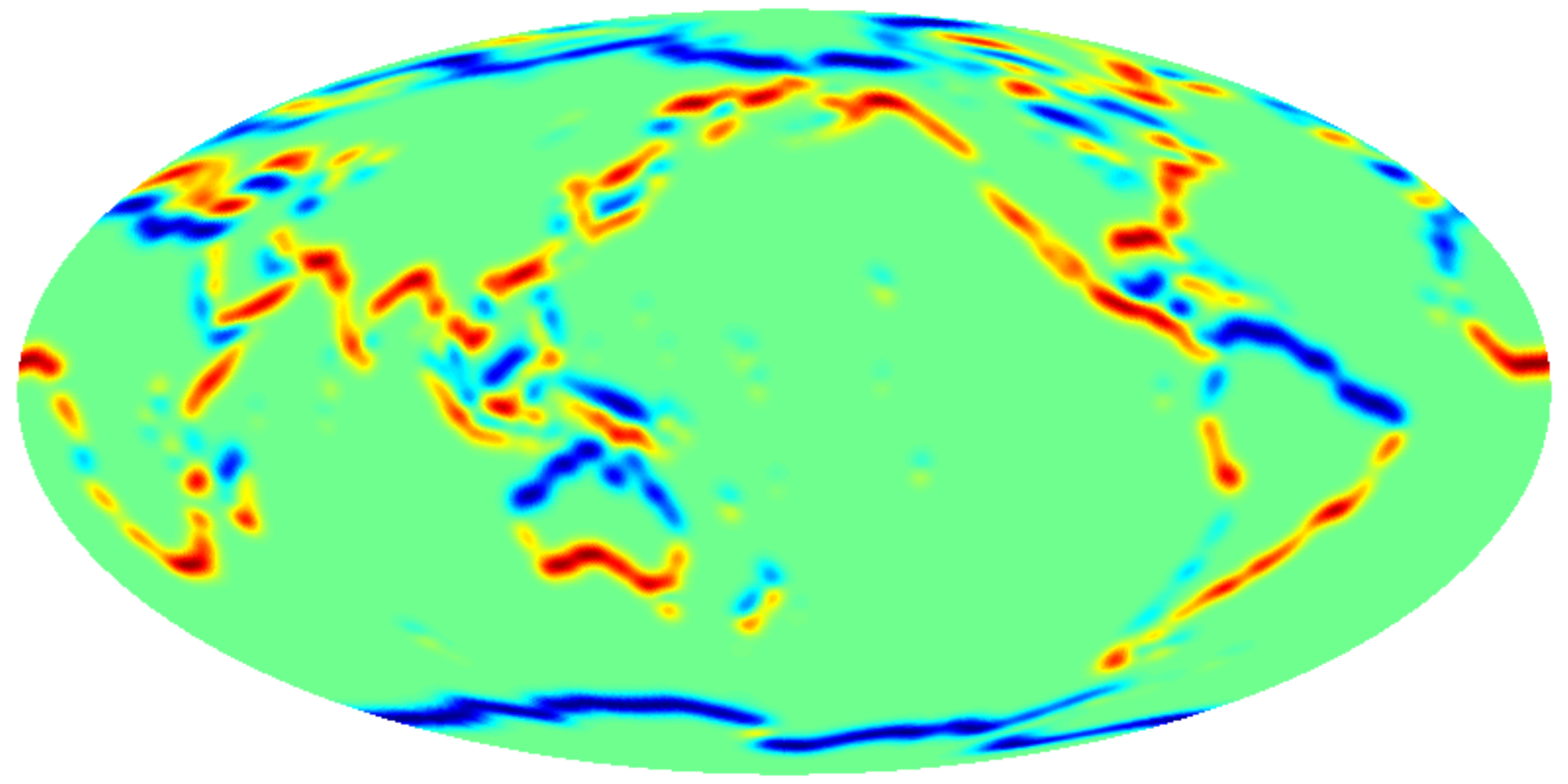}}
  \subfigure[Spherical butterfly wavelet coefficients for scale $\scalea=\scaleb=0.03$ and orientation $\eulc=144^\circ$]{\includegraphics[width=\coeffwidth]{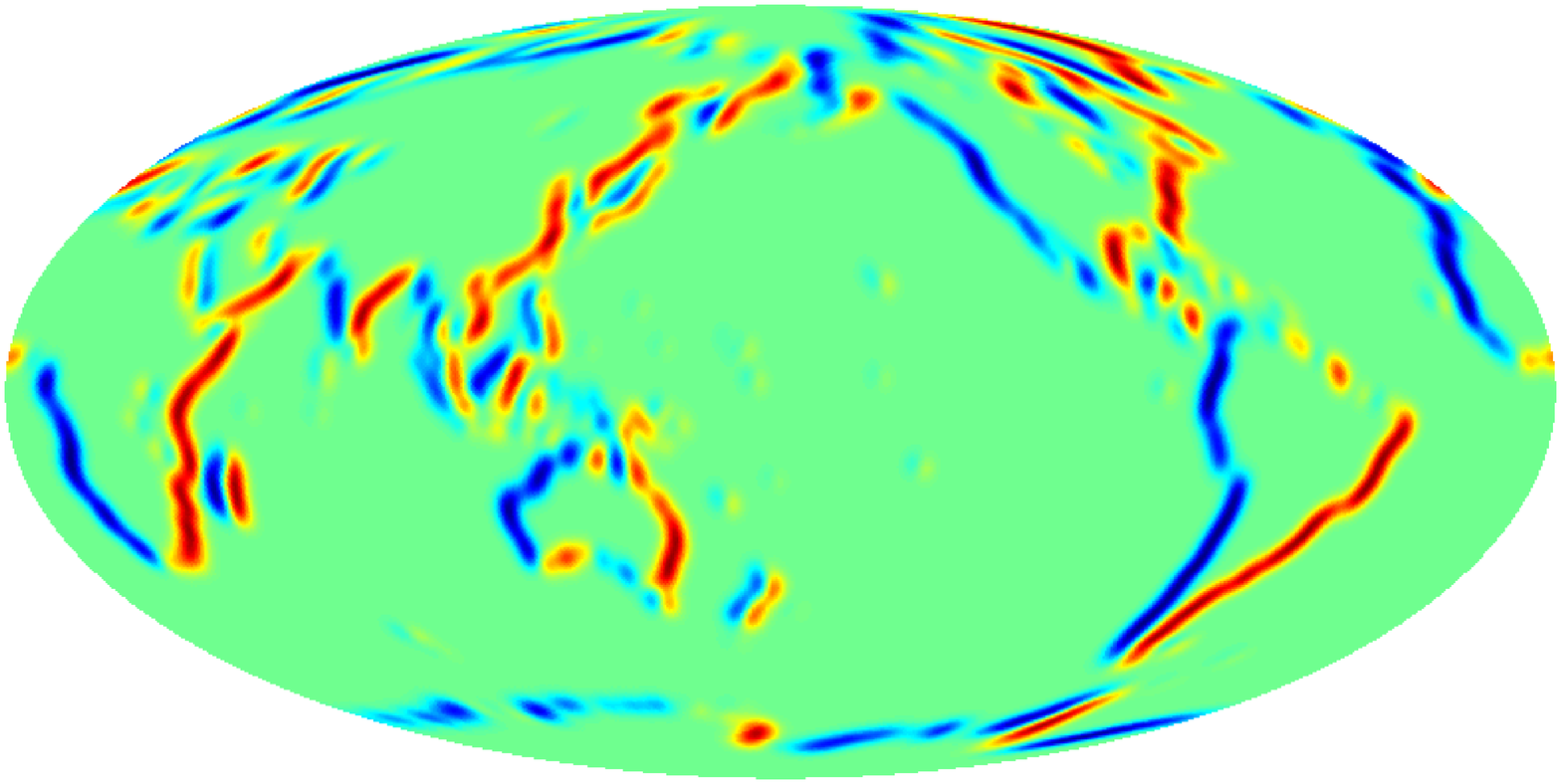}}
  \subfigure[Spherical butterfly wavelet coefficients for scale $\scalea=\scaleb=0.12$ and orientation $\eulc=0^\circ$]{\includegraphics[width=\coeffwidth]{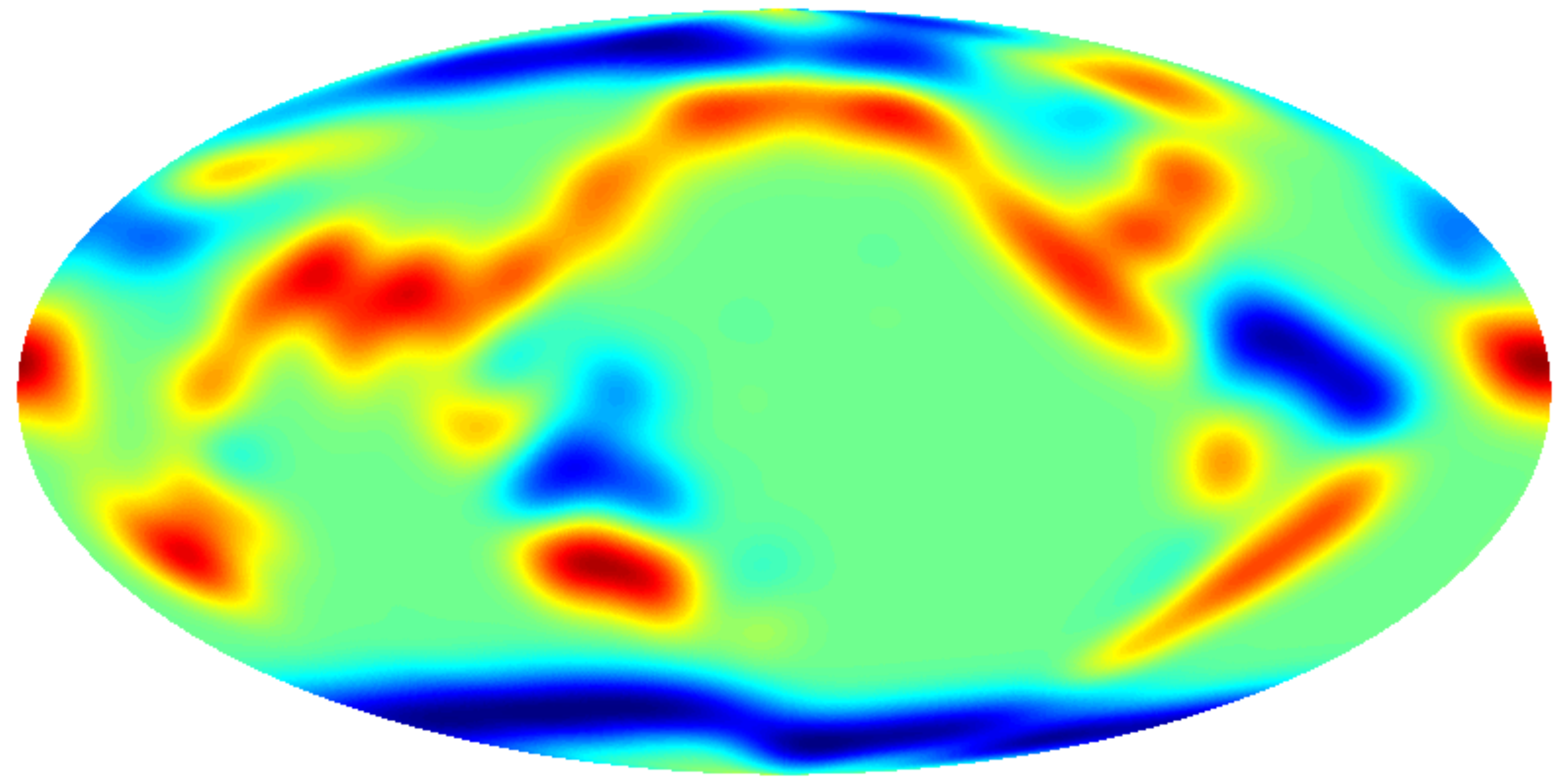}}
  \subfigure[Spherical butterfly wavelet coefficients for scale $\scalea=\scaleb=0.12$ and orientation $\eulc=144^\circ$]{\includegraphics[width=\coeffwidth]{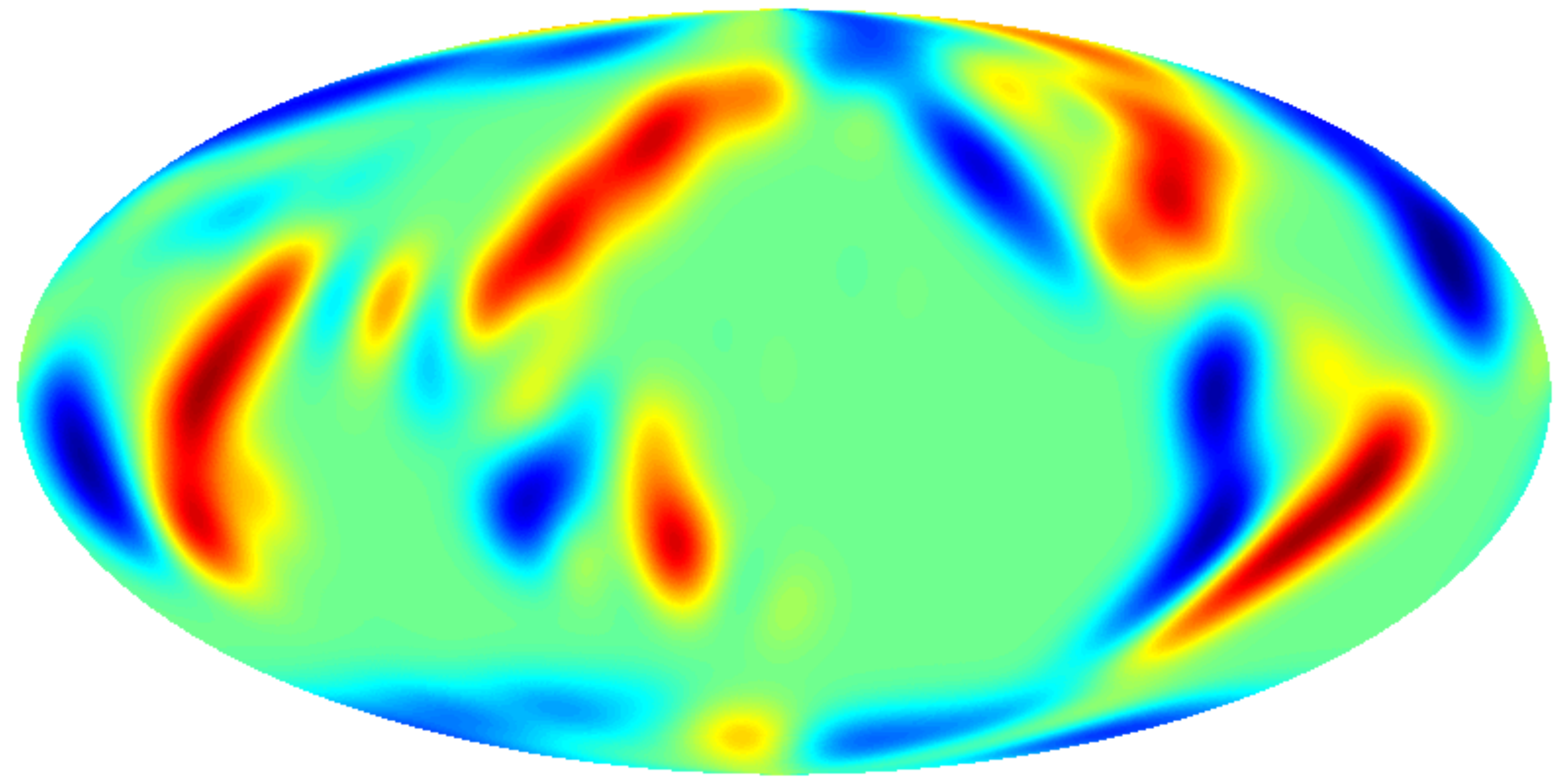}}
\fi 
\caption{Binary Earth data and corresponding spherical butterfly wavelet coefficients.  The wavelet coefficients for combinations of only two scales and two orientations are shown.  Notice how the wavelet coefficient maps corresponding to different oriented wavelets pick out corresponding oriented structure in the data.  As the dilation scale is increased, the scale of the features extracted also increases accordingly.  Directional wavelets therefore allow one not only to probe signal characteristics localised in scale and space, but also in orientation.  However, the concept of orientation on the sphere is necessarily a local one (see text \sectn{\ref{sec:wavelet_transform}} for further details).
}
\label{fig:world}
\end{figure}

We demonstrate in this section the application of our \cswt\ implementation to binary Earth data.  In \fig{\ref{fig:world}} the Earth data and the corresponding spherical wavelet coefficients are shown.  We use the SBW defined in \sectn{\ref{sec:wavelets}} to perform the analysis.  This is a steerable wavelet \cite{wiaux:2005}, however our implementation is in general valid for any directional wavelet.  Notice how the wavelet coefficient maps corresponding to different oriented wavelets pick out corresponding oriented structure in the data.  As the dilation scale is increased, the scale of the features extracted also increases accordingly.

The ability to probe oriented structure in data defined on a sphere is of important practical use.  Certain physical processes may be localised on the sphere in scale, position and orientation (\eg\ signatures of cosmic strings in the CMB \cite{kaiser:1984} or correlations induced in the CMB by the nearby galaxy distribution \cite{crittenden:1996}).  Thus, analysing the statistical properties of spherical wavelet coefficients individually for a range of scales, positions and orientations may allow one to detect such effects with greater significance.
Indeed, using a directional spherical wavelet analysis we have made very strong detections of non-Gaussianity in the CMB \cite{mcewen:2004,mcewen:2006b} and the strongest detection made to date of the ISW effect \cite{mcewen:2006}.

\section{Concluding remarks}
\label{sec:conclusions}

The extension of Euclidean wavelet analysis to the sphere has been
described in the framework presented by \cite{wiaux:2005}, where the
stereographic projection is used to relate the spherical and Euclidean
constructions.  
We extend the concept of the spherical dilation presented
by \cite{wiaux:2005} to anisotropic dilations.  Although anisotropic dilations are of practical use, the resulting basis one projects onto does not classify as a wavelet basis since perfect reconstruction is not possible.


Current and forthcoming data-sets on the sphere, of the CMB for
example, are of considerable size as higher resolutions are
necessary to test new physics.  Consequently, we present fast
algorithms to implement the \cswt\ as an analysis without such
algorithms is not computationally feasible.  A range of
algorithms are described, from the direct quadrature approximation, to
the semi-fast equi-angular algorithm where one rotation is performed
in Fourier space, to the fast azimuthally symmetric and directional algorithms
posed purely in spherical harmonic space.  Posing the \cswt\ purely in
harmonic space naturally ensures the resulting algorithms are tessellation
independent.
The most general fast directional algorithm provides a
saving of $\order(\sqrt{\npix})=\order(\elmax)$ over the direct
implementation and 
may be performed down to a few
arcminutes even with limited computational resources.

Data is observed on a sphere in a range of applications.
In many of these cases the ability to perform a wavelet analysis would
be useful.
For example,
spherical wavelets may be used to probe the CMB for deviations from
the standard assumption of Gaussianity, or to search for compact
objects embedded in the CMB, such as cosmic strings, a predicted but
as yet unobserved phenomenon.  
The extension of wavelet analysis to the sphere enables one to probe
new physics in many areas, and is facilitated on large practical
data-sets by our fast directional \cswt\ algorithm.

\section*{Acknowledgements}

JDM is supported by a Commonwealth (Cambridge) Scholarship from
the Association of Commonwealth Universities and the Cambridge
Commonwealth Trust.
DJM is supported by the UK Particle Physics and Astronomy Research
Council (PPARC).
The implementations described in this paper use the \healpix\
\cite{gorski:2005} and FFTW packages.  
We also acknowledge use of the \yawtb\ Matlab toolbox for the binary Earth data defined on the sphere.

\bibliographystyle{IEEEtran}
\bibliography{fcswt_bibabbrv,fcswt_bib}
%



%

\begin{biography}[{\includegraphics[width=1in,height=1.25in,clip,keepaspectratio]%
{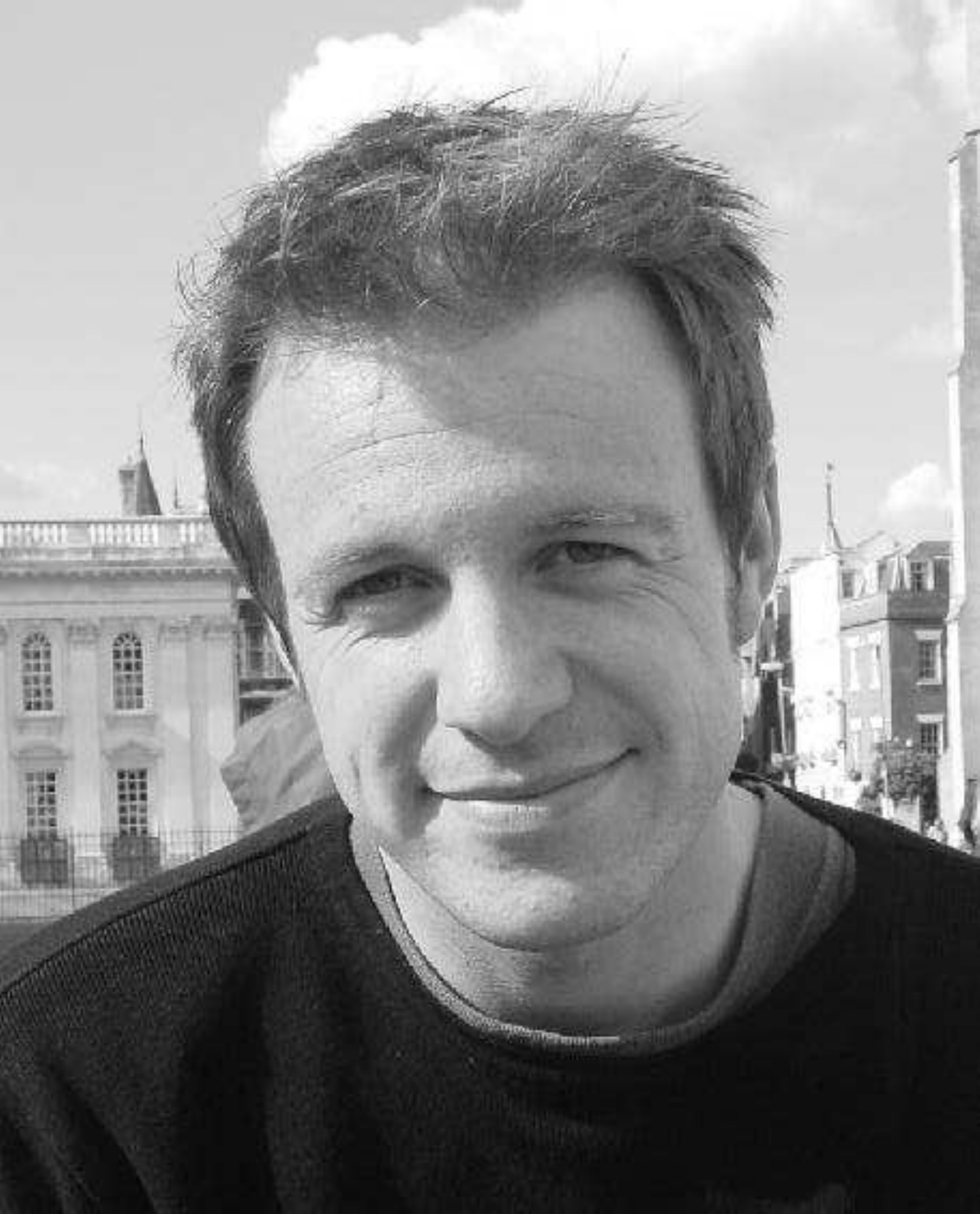}}]{Jason McEwen}
was born in Wellington, New Zealand, in August 1979.
He received a B.E.\ (Hons) degree in Electrical and Computer Engineering
from the University of Canterbury, New Zealand, in 2002.  

Currently, he is working towards a Ph.D.\ degree at the Astrophysics Group,
Cavendish Laboratory, Cambridge.  His area of interests include
spherical wavelets and the cosmic microwave background.
\end{biography}

\begin{biography}[{\includegraphics[width=1in,height=1.25in,clip,keepaspectratio]%
{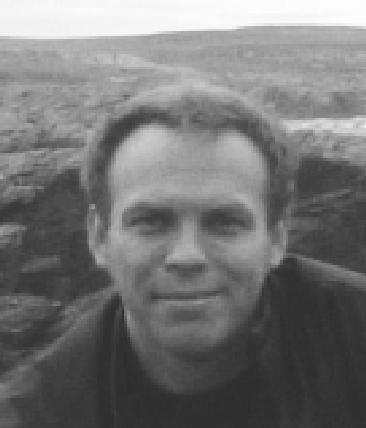}}]{Michael Hobson} 
was born in Birmingham, England, in September 1967. He
received the B.A.\ degree in natural sciences with honours and
the Ph.D.\ degree in astrophysics from the University of
Cambridge, England, in 1989 and 1993 respectively.

Since 1993, he has been a member of the Astrophysics Group of the
Cavendish Laboratory at the University of Cambridge, where he has been
a Reader in Astrophysics and Cosmology since 2003. His research
interests include theoretical and observational cosmology,
particularly anisotropies in the cosmic microwave background,
gravitation, Bayesian analysis techniques and theoretical optics. 
\end{biography}


\begin{biography}[{\includegraphics[width=1in,height=1.25in,clip,keepaspectratio]%
{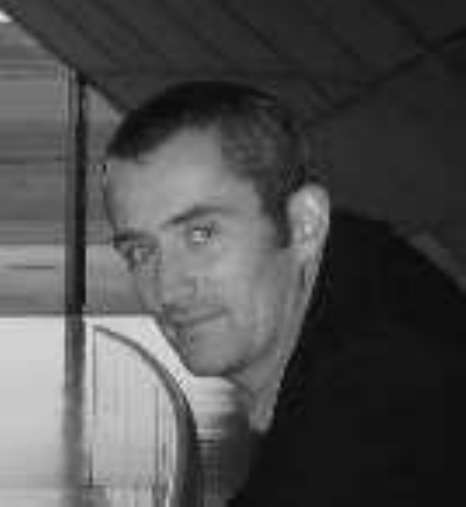}}]{Daniel Mortlock}
was born in Melbourne, Australia, in June 1973.
He received the B.Sc.\ and Ph.D.\ degrees in science from The University
of Melbourne, Melbourne, Australia, in 1994 and 2000, respectively.

From 1999 to 2002 he was research associate at The Cavendish Laboratory,
Cambridge, UK.  In 2002 he moved to The Institute of Astronomy,
Cambridge, UK as a postdoctoral fellow.  His research interests include
cosmology, gravitational lensing and statistical analysis of large data-sets.
\end{biography}

\begin{biography}[{\includegraphics[width=1in,height=1.25in,clip,keepaspectratio]%
{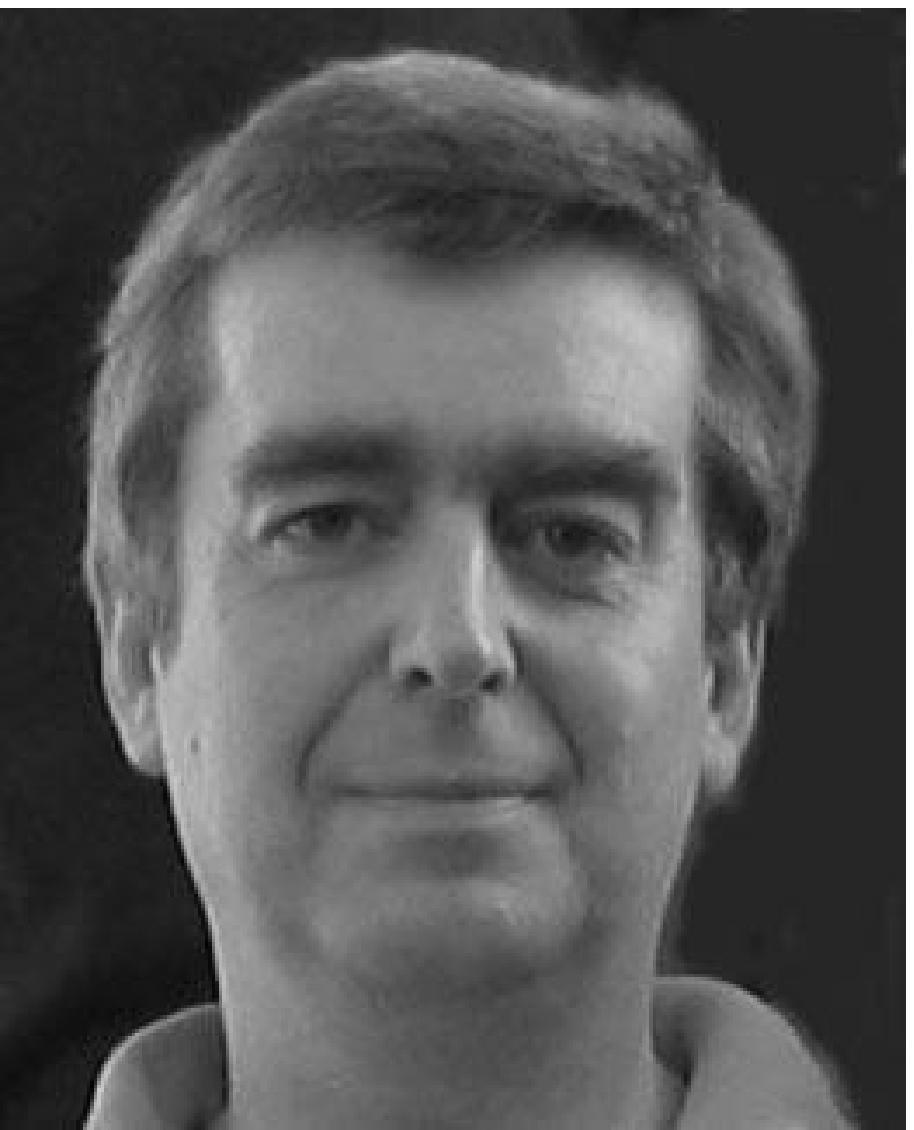}}]{Anthony Lasenby}
was born in Malvern, England, in June 1954. He
received a B.A.\ then M.A.\ from the University of Cambridge in
1975 and 1979, an M.Sc.\ from Queen Mary College, London in 1978
and a Ph.D.\ from the University of Manchester in 1981. 

His Ph.D.\
work was carried out at the Jodrell Bank Radio Observatory
specializing in the Cosmic Microwave Background, which has been a
major subject of his research ever since. After a brief period at
the National Radio Astronomy Observatory in America, he moved
from Manchester to Cambridge in 1984, and has been at the
Cavendish Laboratory Cambridge since then. He is currently Head
of the Astrophysics Group and the Mullard Radio Astronomy
Observatory in the Cavendish Laboratory, and a Deputy Head of the
Laboratory. His other main interests include theoretical physics
and cosmology, the application of new geometric techniques in
computer graphics and electromagnetic modelling, and statistical
techniques in data analysis.
\end{biography}




\end{document}